**Title**

## Tailoring the mechanical properties of 3D microstructures: a deep learning and genetic algorithm inverse optimization framework

**Author list**

Xiao Shang[1], Zhiying Liu[1], Jiahui Zhang[1], Tianyi Lyu[1], Yu Zou[1]*

**Affiliations**

[1]Department of Materials Science and Engineering, University of Toronto, Toronto, ON, Canada, M5S 3E4

*Corresponding author. Email: mse.zou@utoronto.ca

**Abstract**

Materials-by-design has been historically challenging due to complex process-microstructure-property relations. Conventional analytical or simulation-based approaches suffer from low accuracy or long computational time and poor transferability, further limiting their applications in solving the inverse material design problem. Here, we establish a deep learning and genetic algorithm framework that integrates forward prediction and inverse exploration. This framework provides an end-to-end solution to achieve application-specific mechanical properties by microstructure optimization. In this study, we select the widely used Ti-6Al-4V to demonstrate the effectiveness of this framework by tailoring its microstructure and achieving various yield strength and elastic modulus across a large design space, while minimizing the stress concentration factor. Compared with conventional methods, our framework is efficient, versatile, and readily transferrable to other materials and properties. Paired with additive manufacturing's potential in controlling local microstructural features, our method has far-reaching potential for accelerating the development of application-specific, high-performing materials.

## 1. Introduction

Structural materials play a significant role in modern industries, such as personal health care and aerospace technology [1, 2]. The design of new materials, however, is a time-consuming process that could take many decades from research to industrial applications [3]. One reason for such delay is the absence of a clear target application at the time of materials discovery. Teflon, for instance, was discovered as a product of happenstance, spanning almost two decades before its most useful application [4, 5]. One paradigm to address this challenge is "materials-by-design", which emphasises that a comprehensive understanding of the intended application of a new material must be established at the start of a design process [3, 6]. For metallic alloys, a central aspect of materials-by-design is the concept of microstructure, which encompasses heterogenous internal structures such as grain boundaries, grain orientations, and phases. Such heterogeneity of microstructure across multiple length scales controls the mechanical, electrical, magnetic, and optical properties of materials [3]. Consequently, two key components in the materials-by-design paradigm can be identified: (i) forward prediction, which requires robust establishment of microstructure-property (S-P) linkages that accurately predict material properties from their microstructures [7-9]; (ii) inverse exploration, which requires efficient optimization using the established S-P linkages to determine candidate microstructures inversely [6, 10]. To realize these two components, prior studies are classified into two categories: (i) analytical methods [11-16], such as the Voigt and Reuss model, and (ii) numerical methods [17-25], such as finite element analysis (FEA). On the one hand, numerical methods are capable of discretizing materials with arbitrary microstructures and constitutive models, but their high computational power consumption hinders their use in inverse exploration in vast design spaces. On the other hand, analytical based methods enable fast estimations, but are often oversimplified and challenging to establish.

Recent advancement in machine learning (ML) has created new possibilities to address the above challenges in conventional methods. For forward prediction, ML models show advantages in establishing underlying S-P relations [26-28] that are challenging to identify using analytical or simulation-based methods [8]. Moreover, ML models provide fast and accurate surrogate models. Incorporating the ability of the genetic algorithm (GA) to perform optimization in high-dimensional design spaces [29-31], these surrogate models are ideal for solving inverse problems where high throughput exploration is crucial [1, 6, 10, 32].

Among various ML models, one type of deep learning (DL) model, the convolutional neural network (CNN), has gained popularity among materials scientists due to its ability to autonomously learn input features and accurately predict output properties [6, 33-36]. This is particularly favorable when dealing with material S-P relations, because the microstructures of materials are often complex and exhibit high dimensionality [10, 26]. CNN-based models have been proven helpful in forward prediction, solving homogenization problems such as predicting global elastic properties [37] and mechanical responses [38], as well as localization problems such as predicting local stress/strain concentrations [8, 9, 27]. Moreover, their fast inference time makes them well-suited for solving inverse design problems [1, 6, 32]. Current models, however, present limitations in solving real-life materials science problems. They oversimplify materials by using two-dimensional microstructure representations [9, 39], or by ignoring the heterogeneity in material properties caused by local variations of microstructural features [6, 8, 26, 28, 32, 37], such as local elastic modulus differences due to varying grain orientations. Meanwhile, existing work focuses on forward prediction [6, 8, 26-28, 32, 37], and seldomly addresses the more important inverse exploration problem [10]. Moreover, there has not been a DL-powered, comprehensive, end-to-end materials-by-design solution that is efficient and accurately captures real-life material microstructures.

In this study, we address the above limitations by presenting a complete inverse microstructure optimization framework that incorporates both forward prediction and inverse exploration. We demonstrate the effectiveness of this framework in the context of dual-phase Ti-6Al-4V (or Ti64), one of the most widely used titanium alloys in the aerospace and biomedical industries [40, 41]. We demonstrate that with our forward prediction DL models, a high level of accuracy is achieved in the predictions of global stress-strain curves and local stress fields directly from 3D microstructures. The inverse exploration results show how the optimal microstructures are effectively identified based on application-specific mechanical properties using our framework. Collectively, implementing advanced DL models and genetic algorithm, this paper presents a streamlined end-to-end framework that accelerates application-specific material microstructure design.

## 2. Results

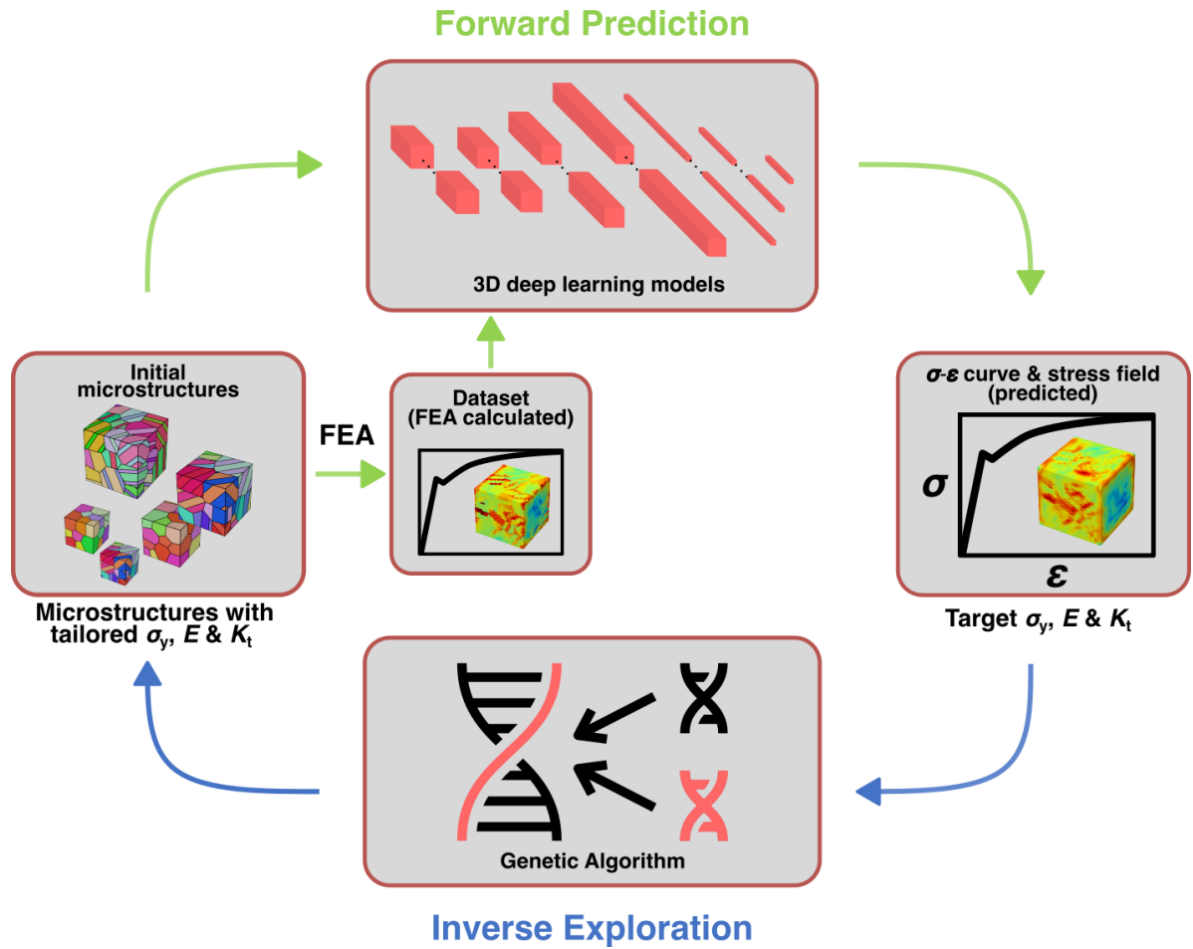

**Figure 1. Illustration of the proposed inverse microstructure optimization framework.** Using a set of microstructure descriptors, including grain morphology, orientation, and phase information, nearly 6000 microstructures are created. FEA is then used to calculate the mechanical responses of microstructures (stress-strain curves and stress fields) under uniaxial tension to create a dataset. This dataset is the input to train 3D deep learning models that perform forward prediction for predicting global stress-strain curves and local stress fields. The predicted stress-strain curves from the deep learning models are then fed into the inverse exploration module, where genetic algorithm is used to inversely identify the optimal microstructures given application-specific optimization objectives.

## 2.1. The inverse microstructure optimization framework

Figure 1 depicts the complete framework, composed of data generation, forward prediction, and inverse exploration. In data generation, a set of three-dimensional Ti64 polycrystalline microstructures embedded with comprehensive microstructural information, including grain location, grain morphology (size and shape), grain orientation, and phase information, are first created. Their mechanical responses, i.e., stress-strain curves and stress fields, are then calculated using finite element analysis (FEA) to generate the full dataset. In forward prediction, the goal is to predict the mechanical responses of the Ti64 microstructures accurately and efficiently. Two DL models, namely a 3D convolutional neural network (3D-CNN) and 3D conditional generative adversarial network (3D-cGAN) are trained on the dataset to address the homogenization and localization problems, respectively. For the homogenization problem, the 3D-CNN model predicts global stress-strain curves from microstructures, while for the localization problem, the 3D-cGAN translates microstructures directly to detailed stress field maps. For inverse exploration, genetic algorithm is used in combination with the deep learning models trained in forward prediction to efficiently search for optimal 3D microstructures given desired mechanical properties, including yield strength ($\sigma_y$), elastic modulus ($E$), and stress concentration factor ($K_t$).

## 2.2. High-fidelity 3D artificial microstructure instantiation

To capture the mechanical behaviours of real-life material microstructures, we need high-fidelity 3D microstructures to develop our models. For this purpose, we generated a series of realistic 3D microstructures that imitate the microstructure of $\alpha+\beta$ Ti-6Al-4V alloys, similar to those described in [42]. A two-level tessellation strategy was utilized to create the microstructures, with level-1 representing equiaxed prior-$\beta$ phase, and level-2 representing $\alpha$ and $\beta$ lamellar (As illustrated in Figure 2A). For more information on the construction of the microstructures, readers are referred to Materials and Methods section and [42]. Nearly 6000 distinct microstructures, with a wide range of geometrical and orientational features, are generated to cover a large data variance in mechanical properties. Geometrically, the number of prior-$\beta$ grains contained within one microstructure ($n_{\text{grains}}$) is varied from 10 to 50 (Figure 2B) and the volume fraction of $\alpha$ phases ($r_\alpha$) is varied from 28% to 80% (Figure 2C). Although typical dual-phase Ti-6Al-4V alloys have 14%-95% $r_\alpha$ [43, 44], 28%-80% $r_\alpha$ was selected for the data generation, because $r_\alpha$ outside this range would result in thin lamellae that are inefficient for the FEA program to mesh and compute. In the 3D microstructures, grain orientations are colour-coded following a modified inverse pole figure (IPF) convention, as demonstrated in Figure 2D. In this colour key, both the $\alpha$ phase and $\beta$ phase are assumed to have cubic symmetry so that one single IPF triangle can be used to show the grain orientations for both phases. Two ideal orientation textures, namely Cube ([011]) and Goss ([001]) orientations, as well as a random orientation are shown as examples in Figure 2D. The three-channel RGB values of each grain colour are converted from the three Euler angles that define the orientation of each grain.

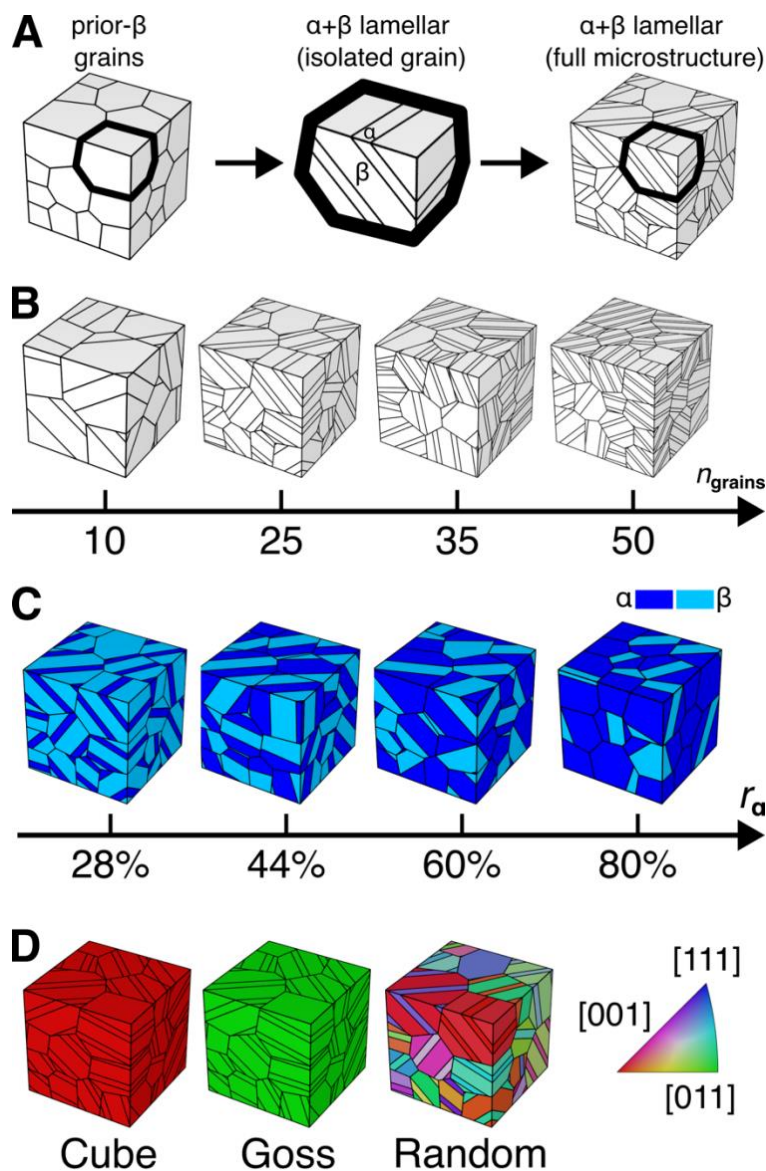

**Figure 2. High-fidelity 3D microstructure instantiation.** (**A**) Illustration of the two-level tessellation. From left to right are equiaxed prior-β grains, α and β lamellar within one isolated prior-β grain, and the final lamellar tessellation. (**B**) Microstructures with varying numbers of prior-β grains, showing four individual microstructures with increasing numbers of prior-β grains ($n_{grains}$) from 10 to 50. The volume fraction of the α phase ($r_α$) is fixed at 28%. (**C**) Microstructures with varying $r_α$, showing examples of four individual microstructures with increasing $r_α$ from 28% to 80%. $n_{grains}$ is fixed at 25. Different colours distinguish α and β phases. (**D**) Microstructures coloured with grain orientations. From left to right are three example microstructures with ideal Cube ([001]) orientation, ideal Goss orientation ([011]), and random orientation. A modified IPF triangle color key is used for colouring.

## 2.3. Forward prediction: microstructure to stress-strain curve using 3D-CNN

Stress-strain curves give valuable information about a material's mechanical properties, such as the yield strength and toughness [45]. Using FEA calculated stress-strain data as ground truth, we trained a 3D-CNN model that successfully predicts the full stress-strain curve of any given microstructure with an $R^2$ score as high as 0.93. The input of this model is rasterized 3D microstructure instances (Supplementary Figures S1C and S3A), and the output is stress values at six distinct points on a stress-strain curve, corresponding to 0.25%, 0.50%, 0.75%, 1.00%, 1.25%, and 1.50% strain. A full stress-strain curve with a total of 1.50% strain can then be reconstructed with these points.

The performance of the 3D-CNN model is shown in Figure 3. When comparing 3D-CNN predicted stress values with FEA ground truth at the six strain points, $R^2$ scores ranging from 0.76 to 0.93 are obtained, with an overall $R^2$ of 0.85 (Figure 3A). Figure 3B and 3C show enhanced comparisons at 1.50% strain (i.e., the plastic regime) with the lowest $R^2$ of 0.76 and at 0.25% strain (i.e., the elastic regime) with the highest $R^2$ of 0.93. The performance of the 3D-CNN model, especially for the plastic regime, can further be improved by using a microstructure input with finer rasterizations, at a cost of more computational power during the model training stage. With finer rasterizations, we find a jump in the $R^2$ scores from 0.76 to 0.82 and from 0.93 to 0.96, at 1.50% and 0.25% strains, respectively. More details on the rasterization of microstructures are discussed in Section 4.1 and Supplementary Figure S1. Next, full stress-strain curves, constructed from FEA calculations and 3D-CNN predictions, are compared in Figure 3D, with three pairs of curves from three randomly selected microstructures in the testing dataset. The curves show remarkable agreement between the FEA simulations and the 3D-CNN model predictions. Although Figure 3D shows a more pronounced discrepancy in the plastic regime (i.e., 0.75%-1.50% strain) than that in the elastic regime (i.e., 0.00%-0.75% strain), such discrepancy in the plastic regime is attributed to the higher stress values in the plastic regime, which lead to larger absolute errors than those in the elastic regime. When the mean absolute percentage error is used, the errors are in fact on a comparable level, i.e., 1.085% vs. 1.172% for the elastic and plastic regimes, respectively (Supplementary Table S1). Moreover, it is worth noting that it takes, on average, 45 minutes on a supercomputer cluster for FEA to complete one microstructure calculation, while with our 3D-CNN model the time is 50 milliseconds on a regular office desktop.

Furthermore, we demonstrated the potential of our 3D-CNN model in real-life applications by comparing its predictions with experimental data. A 3D-XRD reconstructed Ti64 microstructure was used for this purpose, and the stress-strain curves from both the 3D-CNN prediction and experiment are compared. The microstructure (Figure 3E) and the experimental stress-strain curve (Figure 3F) are from the same sample, reproduced with data from [46] and [42], respectively. To rule out potential randomness introduced by the limited number of experimental sample, five new microstructures were generated based on the 3D-XRD sample, by adding randomly distributed noises to the orientations of each grain. The stress-strain curves of these noisy samples are then predicted using the 3D-CNN model. The noises are sampled from a uniform distribution:

$$noise \sim U(Ori_{min}, Ori_{max}) \tag{1}$$

where $Ori_{\text{min}}$ and $Ori_{\text{max}}$ represent the minimum and maximum magnitude of the orientations in the original 3D-XRD microstructure data, respectively, expressed in Euler angles. As demonstrated in Figure 3F, our 3D-CNN model successfully predicts a stress-strain curve that matches well with the experimental data, while the predictions with noisy microstructures exhibit significant deviations from the original experiment data.

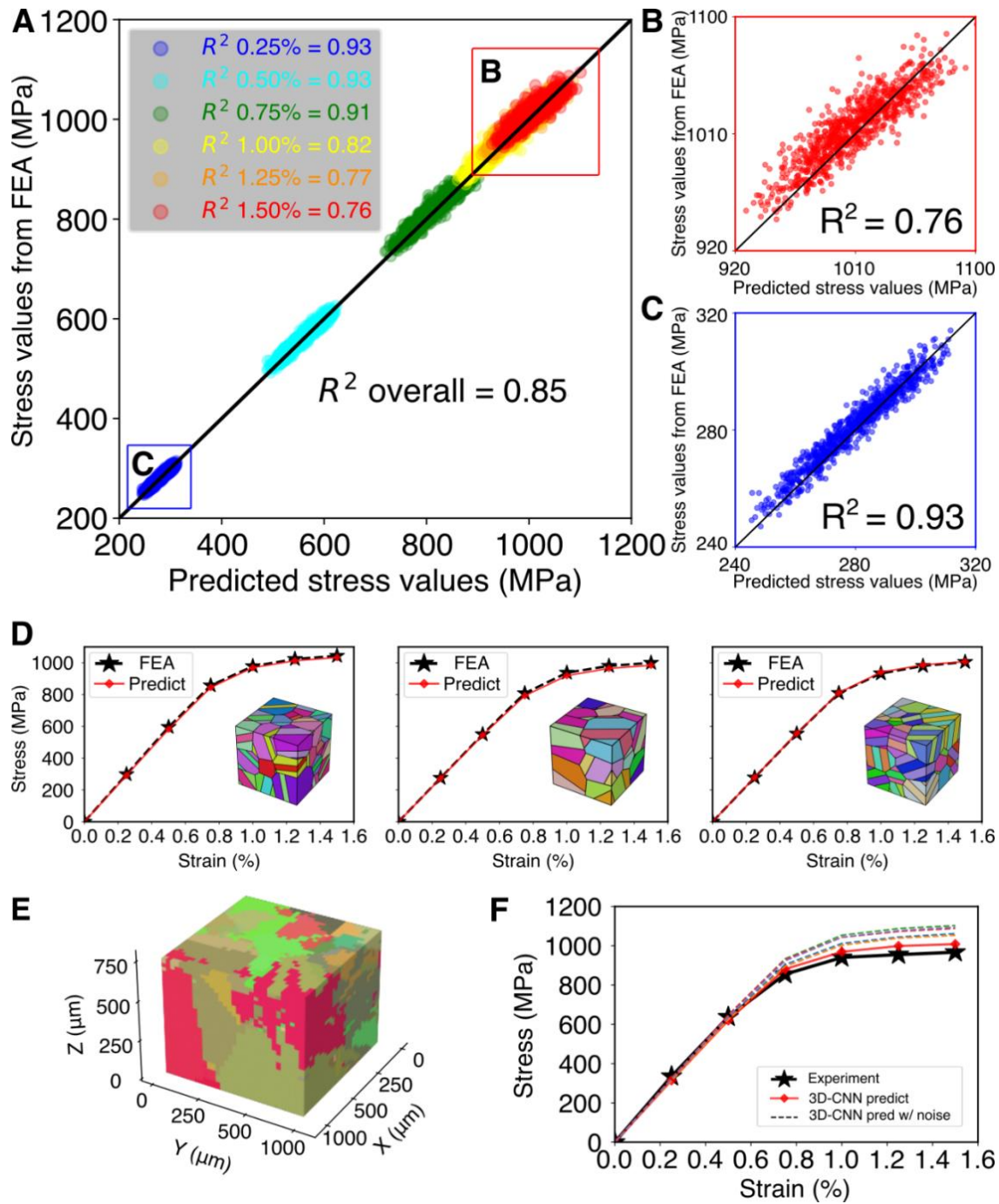

**Figure 3. Performance of the 3D-CNN model.** (**A**) Comparison of 3D-CNN predicted and FEA calculated stress values at the strains of 0.25%, 0.50%, 0.75%, 1.00%, 1.25%, and 1.50% for the entire testing dataset. (**B**) Detailed comparison of 3D-CNN predicted and FEA calculated stress values at 1.50% strain. (**C**) Detailed comparison of 3D-CNN predicted and FEA calculated stress values at 0.25% strain. (**D**) Reconstructed full stress-strain curves obtained from the 3D-CNN predictions and FEA calculations for three randomly sampled microstructures. (**E**) 3D-XRD reconstructed Ti64 microstructure (reproduced using the data from Ref. [46]). Differences in colours signify different grains. (**F**) Comparison of 3D-CNN predicted and experimentally obtained full stress-strain curves for the microstructure in E. The experimentally obtained curve is reproduced using the data from Ref. [42].

## 2.4. Forward prediction: microstructure to stress field using 3D-cGAN

Next, we focus on tackling the localization problem that predicts detailed stress fields and stress concentrations, which are key to failure-related high cycle and low cycle fatigue properties [8, 47-49]. Due to the stringent level of local details required, solving localization problems has proven to be more challenging compared with homogenization problems [8]. To address this challenge, we trained a novel 3D-cGAN model that could efficiently and accurately predict detailed 3D stress fields of a given microstructure under uniaxial tension.

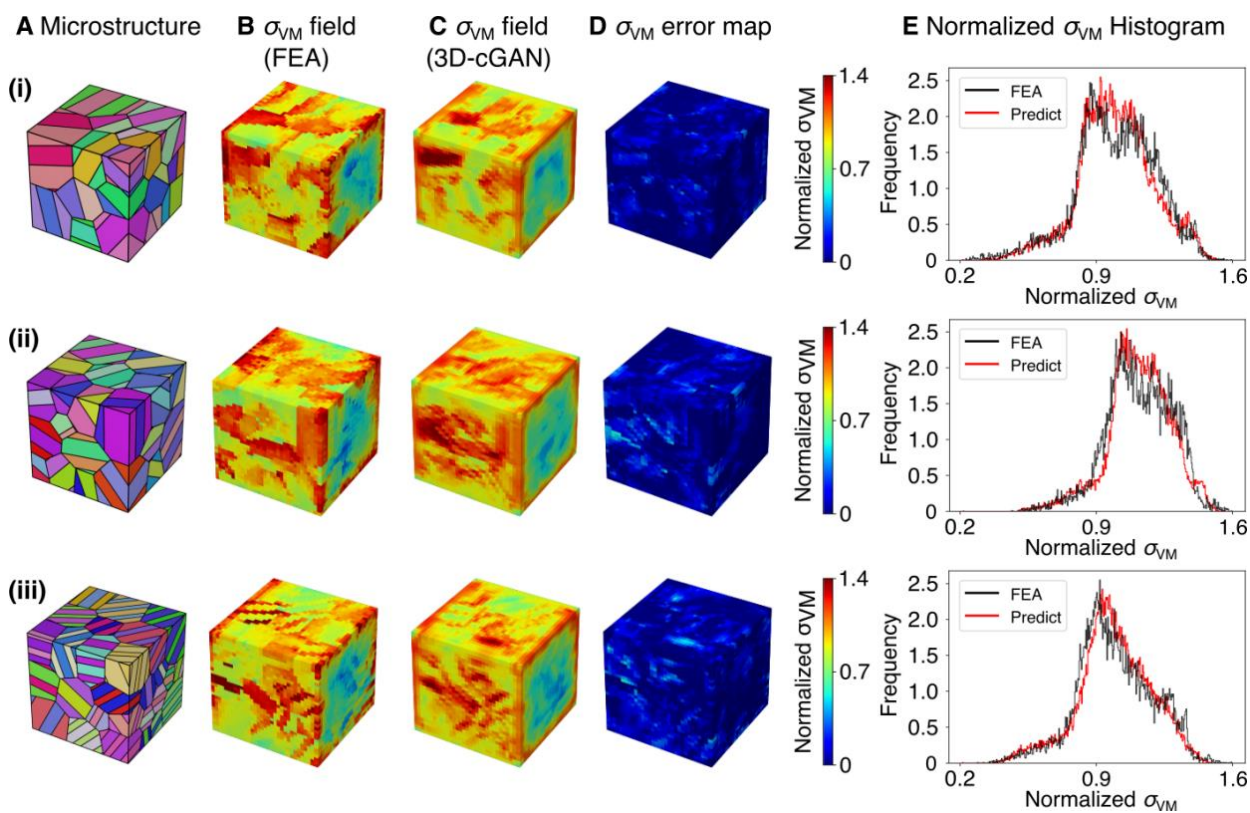

**Figure 4. Performance of the 3D-cGAN model on three randomly selected microstructures.** Columns from left to right are: (**A**) randomly selected 3D microstructures; (**B**) normalized von Mises stress ($\sigma_{VM}$) fields calculated by FEA; (**C**) $\sigma_{VM}$ fields predicted by 3D-cGAN; (**D**) voxel-to-voxel error maps, illustrating the differences between FEA calculated and 3D-cGAN predicted $\sigma_{VM}$; (**E**) $\sigma_{VM}$ histograms of stress distributions from FEA calculation and 3D-cGAN prediction.

The performance of the 3D-cGAN model is given in Figure 4, where we randomly selected three microstructures (Figure 4A) and compared their stress fields from FEA calculations and 3D-cGAN predictions. The von Mises stress ($\sigma_{VM}$) fields are shown in Figure 4B and 4C, corresponding to results calculated from FEA and predicted by the 3D-cGAN model, respectively. Here, the illustrated $\sigma_{VM}$ have been normalized by each microstructure's average stress. The stress fields have a resolution of 32x32x32 voxels, and over 32,000 stress values are used to construct each 3D stress field map. Visual inspection of the results from FEA calculation and 3D-cGAN prediction reveals remarkable accuracy in the stress fields predicted by our 3D-cGAN model. To further quantify the results, the voxel-to-voxel difference in the normalized von Mises stress values between FEA calculation and 3D-cGAN prediction is computed and plotted as 3D $\sigma_{VM}$ error maps (Figure 4D). The errors are predominantly close to zero, suggesting high levels of accuracy in the 3D-cGAN's prediction. Furthermore, histograms of the von Mises values at each voxel on the microstructures are plotted and compared between FEA and 3D-cGAN results (Figure 4E). For each microstructure, the two curves exhibit remarkable agreement, between over 32,000 stress values plotted in each curve.

## 2.5. Inverse exploration: identifying optimal microstructures

The goal of our framework is to identify microstructures with a combination of optimal strength, stiffness, and fatigue performances for given applications. In the inverse exploration, we use our previously trained 3D-CNN model to predict and calculate the yield strength ($\sigma_y$) and elastic modulus ($E$), which are indicators of strength and stiffness performances, respectively. For fatigue properties, we calculate the stress concentration factor ($K_t$) from the stress field predicted by our 3D-cGAN model. Integrating the two DL models with genetic algorithm (GA), we create a multi-objective optimization loop that efficiently performs inverse exploration in the high-dimensional design space of Ti64 microstructures for application-specific mechanical properties.

Figure 5A depicts the inverse exploration optimization loop developed herein. At the start of each optimization generation, GA selects a set of "parent" microstructures to produce new microstructure instances by "crossing over" and "mutating" parameters, including grain morphology, grain location, grain orientation, and the volume fraction of α phase. The previously trained 3D-CNN and 3D-cGAN models are then used to calculate the objective function, which will be used to guide how the next generation of microstructures are selected. This process continues until the stopping criteria is met. In this work, we repeated each optimization task for three times by initializing the optimizations with different random seeds.

To demonstrate the effectiveness of our inverse exploration model, tasks with two optimization targets are performed. The first target ($OPT_{aero}$) is to maximize $\sigma_y$ and $E$ to create strong and stiff Ti64 for aerospace structures, while the second target ($OPT_{med}$) is to maximize $\sigma_y$ and minimize $E$ to create strong Ti64 with biocompatible stiffness for biomedical implants. For both targets, $K_t$ is set to be minimized for a better fatigue performance. Considering the proportional relationship between $\sigma_y$ and $E$, we formulate the $OPT_{aero}$ objective as a scalation multi-objective problem using the traditional GA. For the $OPT_{med}$ problem, which involves a forced trade-off between $\sigma_y$ and $E$, we implemented the NSGA-iii algorithm [50, 51], which is more suited for finding Pareto optimal solutions. More details on the multi-objective optimizations setup are shown in the Materials and Methods section.

As shown in Figure 5B, for both objectives, the simulation runs converged within 40 generations. To further visualize the optimization process and the solution evolution, we project the solutions from one optimization run in $OPT_{aero}$ into a reduced-dimension space using principal components analysis (PCA). The first two principal components, i.e., Principal component 1 and Principal component 2 are used as visualization axis. As shown in Figure 5C, the locations of all the solutions in this optimization run exhibit a clear evolving trend from top-right to middle-left as GA generation evolves. Compared with the first 10 generations, solutions in the last 10 generations show clear convergency in the reduced dimension space.

Finally, we produced a $\sigma_y$ vs $E$ vs $K_t$ chart to compare our inverse exploration results with the microstructures in the original FEA calculated dataset, with the identified optimal microstructures and their stress fields for each task plotted on the sides (Figure 5D).

Detailed optimization results and the distribution of $\sigma_y$ and $E$ in the FEA dataset are also found in Supplementary Figure S2, Tables S2, S3, and S4. Compared with the original dataset, our inverse exploration successfully identified microstructures with desired $\sigma_y$, $E$, and $K_t$ for the various optimization tasks. For OPT$_{aero}$, the optimized microstructures show up to a 4.2% and 2.1% increase in $\sigma_y$ and $E$, respectively, and a 3.2% decrease in $K_t$, compared with the maximum $\sigma_y$ and $E$, and minimum $K_t$ in the original dataset. When comparing the two microstructures with the highest $\sigma_y$ from the dataset and from OPT$_{aero}$, an 8.2% increase in $E$ and an 8.1% decrease in $K_t$ is achieved. For OPT$_{med}$, the optimization converges to a set of nondominated optima, and a Pareto front is identified showing the trade-off between $\sigma_y$ and $E$ (Supplementary Figure S2). To fulfill the requirements for biomedical implants, we selected the microstructure that is the most compliant and has the best fatigue performance to be our optimum, i.e., the minimal $E$ and $K_t$ solution. The microstructure at the dome tip of the Pareto front is also demonstrated, for it provides a good balance between the $\sigma_y$-$E$ trade-off. Other specific optimum can also be selected based on application requirements. Compared with the most compliant microstructure in the original dataset, the minimal $E$ and $K_t$ solution is on average 2.1% more complaint and 1% stronger. The $K_t$ of the minimal $E$ and $K_t$ solution is smaller than 50% of the microstructures in the original dataset as well. The $\sigma_y$-$E$ trade-off solution has smaller $E$ than all microstructures in the original dataset with similar $\sigma_y$. When compared with those having similar $E$ values, the optimal microstructure offers the greatest $\sigma_y$.

Moreover, because the α phase of Ti-6Al-4V generally exhibits higher strength and stiffness than the β phase, we included two more optimization tasks where the influence of the α phase volume fraction ($r_\alpha$) is isolated by being fixed at 50%. These tasks allow us to solely investigate the impact of grain location, morphology, and orientation on the yield strength, elastic modulus, and stress concentration factor of Ti-6Al-4V. The optimization objectives remain invariant as in OPT$_{aero}$ and OPT$_{med}$. The optimization results are shown in Figure 5, Supplementary Figure S2 and Tables S2 and S4. Compared with the microstructures with 50% $r_\alpha$ in the original dataset, the identified optimal microstructures show similar optimality as discussed above. For OPT$_{aero}$, the optimal microstructures show 4.4%, 4.9%, 13.7% improvements in $\sigma_y$, $E$, and $K_t$, respectively. For OPT$_{med}$, the optimal microstructures have $E$ smaller than all microstructures in the original 50% $r_\alpha$ FEA dataset. Meanwhile, the optimal microstructures in OPT$_{med}$ show on average 2.5% and 28.1% improvements in $\sigma_y$ and $K_t$, respectively, compared with the most compliant microstructure in the dataset.

In terms of the optimal microstructures, those from OPT$_{aero}$ have more equiaxial grains with a high α phase fraction averaging 87.2%, while those from OPT$_{med}$ show more and thinner lamellar grains with a low α phase fraction averaging 28.7% (Figure 5D, Supplementary Table S2). When the α phase fraction is fixed at 50%, the microstructures identified in OPT$_{aero}$ and OPT$_{med}$ show on average 13.1%, 14.4%, and 3.8% differences in $\sigma_y$, $E$, and $K_t$, respectively, which are caused solely by the differences in geometries and orientations of the grains.

Furthermore, we performed OPT$_{aero}$ within a reduced design space setting $r_\alpha$ within 28% to 80%, i.e., the $r_\alpha$ range used in our original training data. The results confirm that microstructures with global optimal properties in the given $r_\alpha$ range are successfully discovered (Supplementary Figure S2, and Table S3).

It is worth noting that due to the high-dimensional and complex nature of the microstructure optimization problem, the optimization solution is expected to be non-unique. Despite this, Figure 5B and 5C, as well as the similarities in the identified microstructure in Figure 5D, are indicators that the solutions identified by our design approach are very close to the absolute global optimal.

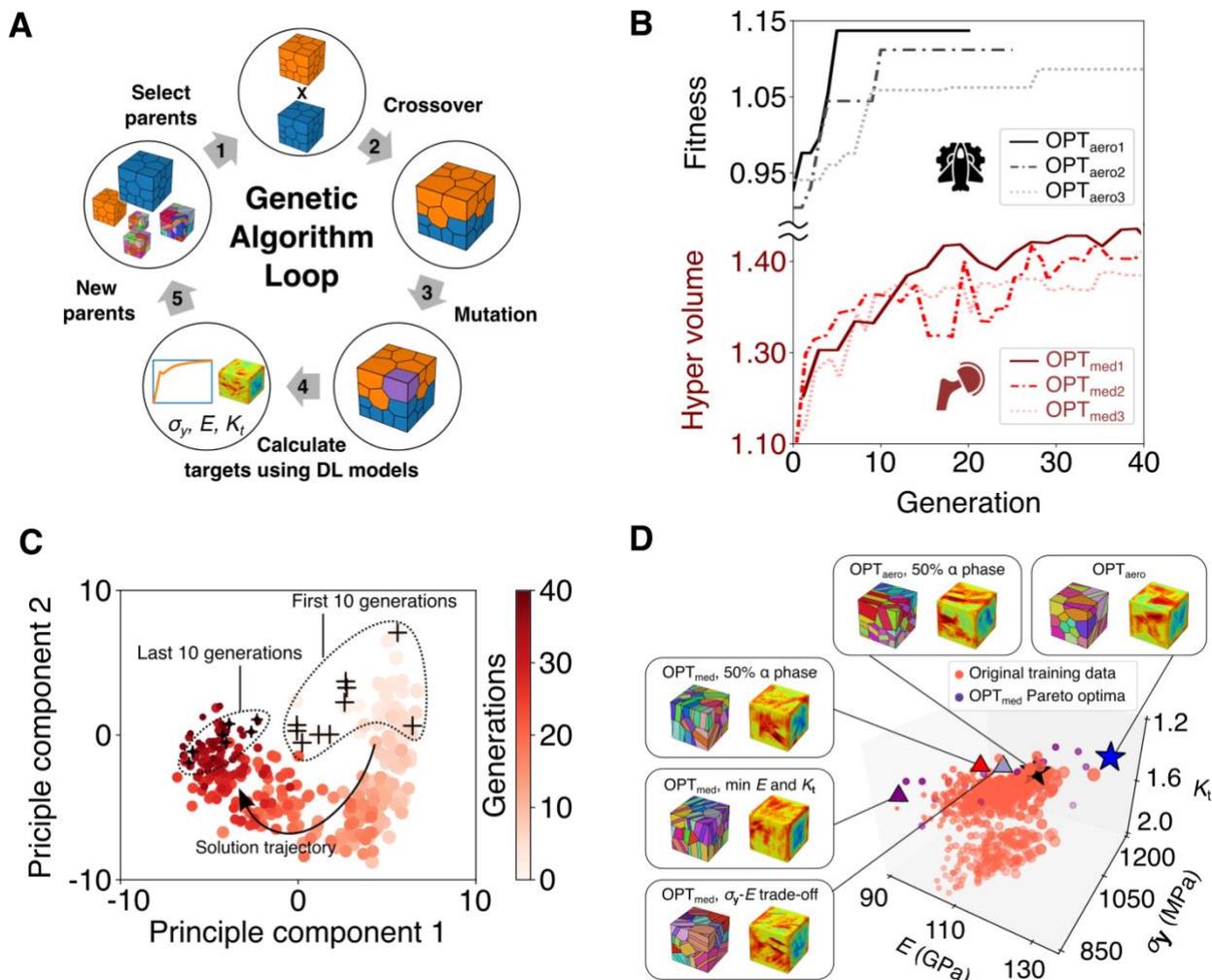

**Figure 5**. **Inverse exploration results.** (**A**) Illustrative genetic algorithm (GA) optimization loop. (**B**) Convergence plot for two optimization objectives OPT$_{aero}$ and OPT$_{med}$. (**C**) Evolution of GA solutions over 40 optimization generations visualized in a reduced dimension space. Solutions from the first and last ten generations are circled and labelled, and the trajectory of solution evolution is marked by a black arrow. (**D**) Comparison of $\sigma_y$, $E$, and $K_t$ between inverse exploration results and original FEA dataset. Identified optimal microstructures and the stress fields for corresponding optimization tasks are visualized. The X, Y and Z axis are elastic modulus ($E$), yield strength ($\sigma_y$), and stress concentration factor ($K_t$), respectively.

## 3. Discussion

In this work, we present an end-to-end inverse microstructure optimization framework integrating the computational efficiency of deep learning (DL) and the optimization capabilities of genetic algorithm (GA). This framework can efficiently identify the optimal microstructures that exhibit application-specific mechanical properties and is readily transferrable to various material systems. Using Ti-6Al-4V alloy as a demonstration, the effectiveness of this framework is demonstrated with two case studies targeting aerospace and biomedical applications. In both cases, the optimal microstructures with desired yield strength and elastic modulus, and low stress concentration factor were identified in less than 8 hours on a desktop computer with an Intel i5-7500 CPU. This could have taken years, if not decades, with a conventional materials design approach (3-5). Hence, the framework proposed in this work is a promising tool in accelerating the design of new material microstructures with targeted material properties.

We also show how the DL models developed herein tackle the stress homogenization and localization problems for materials with complex microstructures. By extracting microstructural features autonomously from 3D microstructures, our DL models are used to predict the mechanical responses of a material microstructure under uniaxial tension, achieving high speed and high accuracy. We first show how our 3D-CNN model can reconstruct full stress-strain curves and, consequently, predict the mechanical properties (e.g., $\sigma_y$ and $E$) of a material in less than 50 milliseconds on a desktop computer. This is compared to over 30 minutes required by conventional FEA methods on supercomputer clusters with 32 parallel CPU cores. The model exhibits high accuracy as measured by its $R^2$ scores as high as 0.93. Using a microstructure reconstructed from 3D-XRD measured data, we demonstrate that our model not only works for artificially generated virtual microstructures but is also capable of capturing the mechanical responses of real-life microstructures. Next, we show that our 3D-cGAN model achieves remarkable accuracy in translating material microstructures directly to stress fields, making it well suited for addressing stress localization problems. Its capability to capture detailed local stress concentrations directly from 3D microstructures opens new doors for addressing fatigue-related material fracture problems.

Furthermore, the proposed framework is highly transferrable to other materials and properties. Following the workflow demonstrated herein in a complex dual-phase Ti64 materials system, one can expect similar or better performance when applying the same workflow to other types of materials, including metals, non-metals, and composite materials. Similarly, this framework can be used towards solving problems in other scientific fields, such as thermal, electromagnetic, acoustic systems etc., given the appropriate training data.

Despite the encouraging results, the authors acknowledge that limitations still exist in this work. Firstly, our current FEA generated dataset introduces idealizations that ignore microstructure defects, such as dislocations and pores. This could be addressed by obtaining more accurate material parameters, as well as using a more advanced FEA model, such as the crystal plasticity module available in Ansys®. Furthermore, because of the transferability of the DL models, this issue could also be addressed by training the models using experiment data. Secondly, manipulating microstructures locally with high precision is challenging in the field of manufacturing. With recent advancement in additive manufacturing [52-54], however, our framework has the potential to guide the design and manufacturing of application-specific materials, thus significantly streamlining the conventional material design process and reducing the time from lab design to industrial applications.

## 4. Materials and methods

### 4.1. High fidelity microstructure generation

The 3D Ti64 microstructures in the dataset were created using in-house developed Bash scripts and Neper [55, 56], an open-source polycrystal generation and meshing software, following a slightly modified strategy described in [42]. Resulting microstructures are dual-phase Ti64 3D images capturing detailed microstructural features, including the grain location, grain morphology, and grain orientations, tessellated in a $1 \times 1 \times 1$ dimensionless cubic domain. The microstructures are preprocessed into rasterized 3D images prior to being used as DL model inputs (Supplementary Figure S3A). For the FEA calculations, we used vectorized microstructure geometries, preventing the potential distortion from rasterization in analyzing stress behaviour during the FEA data generation stage. To further investigate how different levels of rasterization influence the stress behaviours of the 3D-CNN prediction, we performed parametric studies using five different sets of microstructure data rasterized with the resolutions of $16 \times 16 \times 16$, $24 \times 24 \times 24$, $32 \times 32 \times 32$, $40 \times 40 \times 40$, and $48 \times 48 \times 48$, respectively. Supplementary Figure S1A shows the results of the performance of each model. In general, finer resolutions yield better model performances in both the elastic and plastic regimes. By increasing the rasterization size from $16 \times 16 \times 16$ to $48 \times 48 \times 48$, the overall $R^2$ score increases from 0.76 to 0.90 (increase by 17.9%), the $R^2$ score for 0.25% strain from the elastic regime increases from 0.88 to 0.96 (increase by 9.1%), and the $R^2$ score for 1.50% strain from the plastic regime increases from 0.64 to 0.82 (increase by 28.1%). Supplementary Figure S1C shows that the rate of improvement slows down passing $40 \times 40 \times 40$ as the overall $R^2$ score plateaus at around 0.90. As shown in Supplementary Figure S1C, there is a trade-off between computational burden and model accuracy. A good balance between the two is identified at the resolution of around $32 \times 32 \times 32$, which is our selected microstructure rasterization level.

For the generalizability of the trained DL models, the grain orientations of prior-β grains are sampled randomly from the BCC fundamental region. After that, the α grain orientations and lamellar grain plane directions were calculated by enforcing Burger's orientation relations [57]. Supplementary Figure S3B shows the resulting orientation distributions in the fundamental regions of BCC and HCP crystal structures, respectively. Orientation sampling and calculation were done using in-house developed MATLAB code and MTEX [58, 59], an open-source MATLAB toolbox for analyzing and modeling crystallographic textures.

### 4.2. Finite element analysis (FEA)

The mechanical responses data in our dataset were calculated using FEpX [60], a finite element analysis (FEA) software package for modeling global and local mechanical behaviours of polycrystals by employing the elastic-viscoplastic model. FEpX is capable of simulating 1) the mechanical behaviours of polycrystal solids, taking into consideration anisotropic elasticity based on cubic or hexagonal crystal symmetry; 2) anisotropic plasticity, based on rate-dependent slip on a restricted number of systems for cubic or hexagonal symmetry; and 3) evolution of state variables for crystal lattice orientation and slip system strengths. For a more complete description of the kinematics and configuration of the simulation model, the reader is referred to the FEpX theory and methods manual [60].

The FEA simulations were conducted in 3D, simulating the cross section of a tensile test coupon under uniaxial tension with a total strain of 1.5% (Supplementary Figure S4A). The boundary conditions were applied such that one loading face is completely fixed in all degrees of freedoms, while the other loading face is applied with a tensile force at a constant strain rate of 0.001/s along its normal direction. The material properties used for the simulations were obtained from literature [42] and are listed in Supplementary Tables S5 and S6. The simulation calculates reaction forces, which are then converted to engineering stresses, and plotted to reconstruct full stress-strain curves. Six reaction force values at 0.25%, 0.50%, 0.75%, 1.00%, 1.25%, and 1.50% strain were selected to reconstruct the stress-strain curve, so that a smooth curve covering the full elastic regime, as well as the yield point, can be captured while still being computationally efficient in FEA calculations. Generated 3D microstructures are meshed with tetrahedral meshes (Supplementary Figure S1B) in Neper for use in the simulations using an adaptive meshing method. A mesh convergence study was performed to determine the best meshing parameters that balance between computation cost and accuracy. All microstructure generation, meshing, and simulation were performed on supercomputer clusters at the Digital Research Alliance of Canada, using 32 Intel E5-2683 v4 CPUs with 125G memories. Compared with experimental measurements, the crystal plasticity FEA method involves inevitable idealizations. While being minor, the discrepancy due to these idealizations can be more pronounced in the plastic deformation regime compared with that in the elastic regime [61-64]. Moreover, while most material parameters used in our FEA calculations were obtained via experimental methods, some parameters used for calculating the plastic deformations were assumed to limit the magnitude of hardening and suppress the evolution of the saturation strength [42] (Supplementary Table S6). Although these assumptions work generally well when compared with experimental measurements, they contribute to the larger deviations in the plastic deformation regime.

Supplementary Figure S4B shows the distribution of $\sigma_y$ and $E$ throughout the whole dataset, grouped by the number of prior-β grains ($n_{grains}$) and the α phase volume fraction ($r_\alpha$). For each group, the distributions of $\sigma_y$ and $E$ roughly follow a normal distribution and no skewed features have been spotted in the distributions. Overall, 5830 sets of microstructure data were generated, distributed evenly with respect to $n_{grains}$ and $r_\alpha$. The dataset is further divided into a training set and a testing set at an 80/20 split for training and testing the DL models.

### 4.3. Deep learning models

The development and training of our two DL models, i.e., 3D-CNN and 3D-cGAN, are performed using Tensorflow [65], a general framework for deep learning tasks in Python. Originally developed for solving computer vision problems, convolutional neural networks (CNN) [66] are particularly suitable for extracting and learning information from material microstructures due to their ability to learn important local features (e.g., microstructural features) and their relative importance (e.g., interaction between different grains). By extending traditional two-dimensional CNN to three-dimensional, our proposed 3D-CNN model is able to solve the homogenization problem of translating complex 3D microstructures to full stress-strain curves. The detailed model architecture is illustrated and described in Supplementary Figure S5A and Table S7, respectively. Generative adversarial networks (GAN) is a type of DL model that trains a generator and a discriminator based on the game theory [67]. Although generative models suffer from training instability issues, its conditional version, i.e., cGAN, offers better stability using conditional labels during the training of the generator [68]. CGAN presents confirmed capability to translate images pixel-to-pixel [69] and is thus well-suited for solving microstructure stress localization problems where stress fields need to be extracted in a pixel-to-pixel (for 2D) or voxel-to-voxel (for 3D) manner. Furthermore, due to the presence of the discriminator, the selection of cGAN can further facilitate training as it automatically improves the generator's performance without the need for an explicit loss function [68, 69]. In fact, similar cGAN models have been successfully used to study material property-structure relationships [9, 70]. In this work, we extended the original cGAN [69] to a 3D version that translates spatial 3D features to 3D stress maps in a voxel-by-voxel fashion. The model architectures illustration can be found in Supplementary Figure S5B and Table S8. In our study, the generator in the 3D-cGAN model has an all-convolutional modified U-Net [71] architecture. While the addition of max-pooling layers may offer better efficiency in data usage and model training than strided convolutional layers, all-convolutional networks have been proven to work well in image-to-image translation tasks [69, 72, 73]. To further justify our selection, we compared three generators with i) all-convolutional layers, ii) alternating convolutional and max-pooling layers, and iii) all-max-pooling layers. The detailed model architectures and numbers of trainable parameters are shown in Supplementary Tables S8 and S9. The Supplementary Figure S6 compares the performances of the generators. Out of the three models, our all-convolutional architecture outperforms the other models in terms of prediction accuracy. The all-max-pooling architecture failed to capture localized stress concentrations, despite offering the most efficient training process due to its least numbers of trainable parameters (Supplementary Tables S8 and S9). Although the generator with alternating convolutional and max-pooling layers seems to offer a good balance between training efficiency and model accuracy, in this work we implement transfer learning to improve model training efficiency of our all-convolutional generator. Our 3D-CNN model is used as a pre-trained model for our 3D-cGAN generator (Supplementary Table S8), reducing its trainable parameters to only 859,265, which outperforms the generator with alternating convolution and max-pooling layers. Compared with when no transfer learning is implemented, this strategy reduced the training time by over 15%.

All DL models are trained on supercomputer clusters available at the Digital Research Alliance of Canada, using a single NVIDIA V100 GPU with 32G memories. Hyperparameter tuning was performed to fine-tune model performance.

### 4.4. Genetic algorithm optimization

An in-house Python module was developed for performing the genetic algorithm (GA) optimization, integrating: 1) MATLAB for grain orientation calculation, 2) Neper for 3D microstructure generation, 3) 3D-CNN and 3D-cGAN models for mechanical property inference, and 4) PyGAD [74] and Pymoo [75], open-source Python implementations for genetic algorithm (GA) execution. These Python modules provide an intuitive end-to-end package to search for optimal microstructures given application-specific targets.

The two optimization objective functions demonstrated in this study are:

$$OPT_{aero}: \textbf{max } sum\left(\sigma_y(\boldsymbol{\theta}, \boldsymbol{X}, r_\alpha) + E(\boldsymbol{\theta}, \boldsymbol{X}, r_\alpha) - K_t(\boldsymbol{\theta}, \boldsymbol{X}, r_\alpha)\right) \tag{2}$$

$$OPT_{med}: \textbf{min } \frac{1}{\sigma_y(\boldsymbol{\theta}, \boldsymbol{X}, r_\alpha)}, E(\boldsymbol{\theta}, \boldsymbol{X}, r_\alpha), K_t(\boldsymbol{\theta}, \boldsymbol{X}, r_\alpha) \tag{3}$$

$$\textbf{sbj.t.}: \boldsymbol{\theta} \in (0,360) \text{ or } (0,180) \tag{4}$$

$$\boldsymbol{X} \in \frac{1}{32} \times \mathbb{N}(1,32) \tag{5}$$

$$r_\alpha \in (0.14, 0.95) \tag{6}$$

where $\boldsymbol{\theta}$ are the 25 grain orientations each expressed as a set of Euler angles $\alpha$, $\beta$, and $\Upsilon$ ($\alpha \in (0,360)$, $\beta \in (0,180)$, and $\Upsilon \in (0,360)$) in degrees; $\boldsymbol{X}$ are the centres of the 25 grains expressed as discrete coordinates ($x$, $y$, $z$) in a 3D Cartesian system, and are only allowed to be at the centres of the $32 \times 32 \times 32$ voxels; $r_\alpha$ is the volume fraction of the $\alpha$ phase for a given microstructure, and it is set to vary between 0.14-0.95 to match those normally seen in Ti-6Al-4V alloys [43, 44]. $\sigma_y$ and $E$ are the yield strength and the elastic modulus, respectively. They are derived from the reconstructed full stress-strain curves, where $\sigma_y$ is identified at 0.2% offset, and $E$ is the slope of the elastic regime in a stress-strain curve (Supplementary Figure S4A). $K_t$ is the stress concentration factor, which is calculated using:

$$K_t = \frac{\sigma_{max}}{\sigma_{ave}} \tag{7}$$

where $\sigma_{max}$ is the maximum stress value of the whole stress field, and $\sigma_{ave}$ is the average stress of the plane where $\sigma_{max}$ is found, which is normal to the loading direction Y (Supplementary Figure S4C). The values of $\sigma_y$ and $E$ are scaled down to unitless numbers within the optimization loops to ensure $\sigma_y$, $E$, and $K_t$ are on a comparable scale, i.e., within the range of 0.8 to 2.1. For convergency indication, in the traditional GA loops the fitness is calculated, and in the NSGA-iii loops the hypervolume indicator [76] is used.

Due to the limitations of the FEA model implemented in this work, microstructural defects, such as dislocations and pores, are not considered. Thus, as shown in Supplementary Figure S4B, $n_{grains}$ have limited influence on how $\sigma_y$ and $E$ are distributed in our dataset. As a result, only the microstructures with $n_{grains} = 25$ are considered in the optimization tasks demonstrated herein. In the GA loop, optimization parameters $\boldsymbol{\theta}$, $\boldsymbol{X}$, and $r_\alpha$ are referred to as "genes". A combination of genes is used to determine a 3D microstructure, or a "chromosome", and multiple chromosomes are grouped as a "population", which undergo GA cycles iteratively as depicted in Figure 5A. In our GA loop, one 3D microstructure (chromosome) consists of 151 genes, one population consists of four 3D microstructures (chromosomes) for single-point crossover, and the mutation rate was set to 10%. For each set of optimization objectives, the optimization was repeated three times initialized with different random seeds.

**Data availability**

Data will be made available on reasonable request. The complete framework has been released as open source codes in a Github repository at: https://github.com/xshang93/MsInverseDesign.

**Acknowledgments**

This work was supported by the Natural Sciences and Engineering Research Council of Canada (NSERC) Discovery Grant (RGPIN-2018-05731), Centre for Analytics and Artificial Intelligence Engineering (CARTE) Seed Funding program, New Frontiers in Research Fund-Exploration (NFRFE-2019-00603), and the Data Sciences Institute Catalyst Grant. We acknowledge the Digital Research Alliance of Canada for providing computational resources. We also acknowledge Dr. Romain Quey at CNRS and Mines Saint-Etienne and Dr. Matthew Kasemer at the University of Alabama for their generous help in setting up the FEA simulations, and Dr. Jason Hattrick-Simpers and Dr. Kangming Li at the University of Toronto for their helpful discussions in machine learning. X.S. acknowledges the financial support from the Ontario Graduate Scholarship (OGS) and the NSERC Canada Graduate Doctoral Scholarship (CGS-D).

**CRediT authorship contribution statement:**

**Xiao Shang:** Conceptualization, Methodology, Writing – Original Draft, Writing – Review & Editing. **Yu Zou:** Conceptualization, Supervision, Writing – Review & Editing. **Zhiying Liu:** Methodology, Writing – Review & Editing. **Jiahui Zhang:** Methodology, Writing – Review & Editing. **Tianyi Lyu:** Methodology, Writing – Review & Editing.

**Declaration of Interests:** The authors declare that they have no known competing financial interests or personal relationships that could have appeared to influence the work reported in this paper.

# References

[1] Sanchez-Lengeling, B. and A. Aspuru-Guzik, *Inverse molecular design using machine learning: Generative models for matter engineering.* Science, 2018. **361**(6400): p. 360-365. https://doi.org/10.1126/science.aat2663.

[2] Segal, D., *Materials for the 21st Century.* 2017: Oxford University Press.

[3] Sutton, A.P. and A.P. Sutton, *Materials by design*, in *Concepts of Materials Science*. 2021, Oxford University Press. p. 102-113.

[4] Pyzer-Knapp, E.O., et al., *What Is High-Throughput Virtual Screening? A Perspective from Organic Materials Discovery.* Annu. Rev. Mater. Res., 2015. **45**(1): p. 195-216. https://doi.org/10.1146/annurev-matsci-070214-020823.

[5] APSNews, *Discovery of Teflon.* 2021, APS News.

[6] Yang, Z., et al., *Deep learning approaches for mining structure-property linkages in high contrast composites from simulation datasets.* Comput. Mater. Sci., 2018. **151**: p. 278-287. https://doi.org/10.1016/j.commatsci.2018.05.014.

[7] Cecen, A., et al., *Material structure-property linkages using three-dimensional convolutional neural networks.* Acta Mater., 2018. **146**: p. 76-84. https://doi.org/10.1016/j.actamat.2017.11.053.

[8] Yang, Z., et al., *Establishing structure-property localization linkages for elastic deformation of three-dimensional high contrast composites using deep learning approaches.* Acta Mater., 2019. **166**: p. 335-345. https://doi.org/10.1016/j.actamat.2018.12.045.

[9] Yang, Z., C.-H. Yu, and M.J. Buehler, *Deep learning model to predict complex stress and strain fields in hierarchical composites.* Sci. Adv., 2021. **7**(15): p. eabd7416. https://www.science.org/doi/10.1126/sciadv.abd7416.

[10] Paulson, N.H., et al., *Reduced-order structure-property linkages for polycrystalline microstructures based on 2-point statistics.* Acta Mater., 2017. **129**: p. 428-438. https://doi.org/10.1016/j.actamat.2017.03.009.

[11] Reuß, A., *Berechnung der fließgrenze von mischkristallen auf grund der plastizitätsbedingung für einkristalle.* ZAMM-Journal of Applied Mathematics and Mechanics/Zeitschrift für Angewandte Mathematik und Mechanik, 1929. **9**(1): p. 49-58.

[12] Voigt, W., *Ueber die Beziehung zwischen den beiden Elasticitätsconstanten isotroper Körper.* Ann. Phys. (Berlin), 1889. **274**(12): p. 573-587. https://doi.org/10.1002/andp.18892741206.

[13] Aboudi, J., *The generalized method of cells and high-fidelity generalized method of cells micromechanical models - A review.* Mech. Adv. Mater. Struc., 2004. **11**(4-5): p. 329-366. https://doi.org/10.1080/15376490490451543.

[14] Berveiller, M. and A. Zaoui, *An extension of the self-consistent scheme to plastically-flowing polycrystals.* J. Mech. Phys. Solids, 1978. **26**(5): p. 325-344. https://doi.org/10.1016/0022-5096(78)90003-0.

[15] Hashin, Z., *Assessment of the Self Consistent Scheme Approximation: Conductivity of Particulate Composites.* J. Comps. Mater., 1968. **2**(3): p. 284-300. https://doi.org/10.1177/002199836800200302.

[16] Mori, T. and K. Tanaka, *Average stress in matrix and average elastic energy of materials with misfitting inclusions.* Acta Metall., 1973. **21**(5): p. 571-574. https://doi.org/10.1016/0001-6160(73)90064-3.

[17] Eisenlohr, P., et al., *A spectral method solution to crystal elasto-viscoplasticity at finite strains.* Int. J. Plasticity., 2013. **46**: p. 37-53. https://doi.org/10.1016/j.ijplas.2012.09.012.

[18] Feyel, F., *Multiscale FE2 elastoviscoplastic analysis of composite structures.* Comput. Mater. Sci., 1999. **16**(1): p. 344-354. https://doi.org/10.1016/S0927-0256(99)00077-4.

[19] Feyel, F. and J.-L. Chaboche, *FE2 multiscale approach for modelling the elastoviscoplastic behaviour of long fibre SiC/Ti composite materials.* Comput. Method App. M., 2000. **183**(3): p. 309-330. https://doi.org/10.1016/S0045-7825(99)00224-8.

[20] Kamiński, M., *Boundary element method homogenization of the periodic linear elastic fiber composites.* Eng. Anal. Bound. Elem., 1999. **23**(9): p. 815-823. https://doi.org/10.1016/S0955-7997(99)00029-6.

[21] Lee, S.B., R.A. Lebensohn, and A.D. Rollett, *Modeling the viscoplastic micromechanical response of two-phase materials using Fast Fourier Transforms.* Int. J. Plasticity., 2011. **27**(5): p. 707-727. https://doi.org/10.1016/j.ijplas.2010.09.002.

[22] Miehe, C., J. Schotte, and M. Lambrecht, *Homogenization of inelastic solid materials at finite strains based on incremental minimization principles. Application to the texture analysis of polycrystals.* J. Mech. Phys. Solids, 2002. **50**(10): p. 2123-2167. https://doi.org/10.1016/S0022-5096(02)00016-9.

[23] Okada, H., Y. Fukui, and N. Kumazawa, *Homogenization method for heterogeneous material based on boundary element method.* Comput. Struct., 2001. **79**(20): p. 1987-2007. https://doi.org/10.1016/S0045-7949(01)00121-3.

[24] Smit, R.J.M., W.A.M. Brekelmans, and H.E.H. Meijer, *Prediction of the mechanical behavior of nonlinear heterogeneous systems by multi-level finite element modeling.* Comput. Method App. M., 1998. **155**(1): p. 181-192. https://doi.org/10.1016/S0045-7825(97)00139-4.

[25] Terada, K. and N. Kikuchi, *A class of general algorithms for multi-scale analyses of heterogeneous media.* Comput. Method App. M., 2001. **190**(40): p. 5427-5464. https://doi.org/10.1016/S0045-7825(01)00179-7.

[26] Jung, J., et al., *An efficient machine learning approach to establish structure-property linkages.* Comput. Mater. Sci., 2019. **156**: p. 17-25. https://doi.org/10.1016/j.commatsci.2018.09.034.

[27] Herriott, C. and A.D. Spear, *Predicting microstructure-dependent mechanical properties in additively manufactured metals with machine- and deep-learning methods.* Comput. Mater. Sci., 2020. **175**: p. 109599. https://doi.org/10.1016/j.commatsci.2020.109599.

[28] Liu, R., et al., *Machine learning approaches for elastic localization linkages in high-contrast composite materials.* Integr. Mater. Manuf. I., 2015. **4**(1): p. 192-208. https://doi.org/10.1186/s40192-015-0042-z.

[29] Paszkowicz, W., *Genetic Algorithms, a Nature-Inspired Tool: Survey of Applications in Materials Science and Related Fields.* Mater. Manuf. Process., 2009. **24**(2): p. 174-197. 10.1080/10426910802612270.

[30] Bhoskar, M.T., et al., *Genetic Algorithm and its Applications to Mechanical Engineering: A Review.* Mater. Today-Proc., 2015. **2**(4): p. 2624-2630. https://doi.org/10.1016/j.matpr.2015.07.219.

[31] Paszkowicz, W., *Genetic Algorithms, a Nature-Inspired Tool: A Survey of Applications in Materials Science and Related Fields: Part II.* Mater. Manuf. Process., 2013. **28**(7): p. 708-725. https://doi.org/10.1080/10426914.2012.746707.

[32] Liu, R., et al., *Context Aware Machine Learning Approaches for Modeling Elastic Localization in Three-Dimensional Composite Microstructures.* Integr. Mater. Manuf. I., 2017. **6**(2): p. 160-171. https://doi.org/10.1007/s40192-017-0094-3.

[33] Cang, R., et al., *Microstructure representation and reconstruction of heterogeneous materials via deep belief network for computational material design.* J. Mech. Design, 2017. **139**(7): p. 071404.

[34] Li, X., et al., *A transfer learning approach for microstructure reconstruction and structure-property predictions.* Sci. Rep., 2018. **8**(1): p. 13461.

[35] Liu, R., et al. *Materials discovery: Understanding polycrystals from large-scale electron patterns*. in *2016 IEEE International Conference on Big Data (Big Data)*. 2016. IEEE.

[36] Liu, R., et al. *Deep learning for chemical compound stability prediction*. in *Proceedings of ACM SIGKDD workshop on large-scale deep learning for data mining (DL-KDD)*. 2016.

[37] Rao, C. and Y. Liu, *Three-dimensional convolutional neural network (3D-CNN) for heterogeneous material homogenization*. Comput. Mater. Sci., 2020. **184**: p. 109850. https://doi.org/10.1016/j.commatsci.2020.109850.

[38] Frankel, A.L., et al., *Predicting the mechanical response of oligocrystals with deep learning*. Comput. Mater. Sci., 2019. **169**: p. 109099. https://doi.org/10.1016/j.commatsci.2019.109099.

[39] Fast, T. and S.R. Kalidindi, *Formulation and calibration of higher-order elastic localization relationships using the MKS approach*. Acta Mater., 2011. **59**(11): p. 4595-4605. https://doi.org/10.1016/j.actamat.2011.04.005.

[40] Liu, Z., et al., *A Review on Additive Manufacturing of Titanium Alloys for Aerospace Applications: Directed Energy Deposition and Beyond Ti-6Al-4V*. JOM, 2021. **73**(6): p. 1804-1818. https://doi.org/10.1007/s11837-021-04670-6.

[41] Liu, Z., et al., *High-speed nanoindentation mapping of a near-alpha titanium alloy made by additive manufacturing*. J. Mater. Res., 2021. **36**(11): p. 2223-2234. 10.1557/s43578-021-00204-7.

[42] Kasemer, M., R. Quey, and P. Dawson, *The influence of mechanical constraints introduced by β annealed microstructures on the yield strength and ductility of Ti-6Al-4V*. J. Mech. Phys. Solids, 2017. **103**: p. 179-198. https://doi.org/10.1016/j.jmps.2017.03.013.

[43] Ren, Y., et al. *Influence of primary α-phase volume fraction on the mechanical properties of Ti-6Al-4V alloy at different strain rates and temperatures*. in *IOP Conf. Ser.: Mater. Sci. Eng.* 2018. IOP Publishing.

[44] Villa, M., et al., *Microstructural modeling of the α+ β phase in Ti-6Al-4V: a diffusion-based approach*. Metall. Mater. Trans. A, 2019. **50**: p. 2898-2911.

[45] Yang, C., et al., *Prediction of composite microstructure stress-strain curves using convolutional neural networks*. Mater. Design, 2020. **189**: p. 108509. https://doi.org/10.1016/j.matdes.2020.108509.

[46] Wielewski, E., et al., *Three-dimensional α colony characterization and prior-β grain reconstruction of a lamellar Ti–6Al–4V specimen using near-field high-energy X-ray diffraction microscopy*. J. Appl. Crystallogr., 2015. **48**(4): p. 1165-1171.

[47] Priddy, M.W., et al., *Strategies for rapid parametric assessment of microstructure-sensitive fatigue for HCP polycrystals*. Int. J. Fatigue, 2017. **104**: p. 231-242. https://doi.org/10.1016/j.ijfatigue.2017.07.015.

[48] Przybyla, C., et al., *Microstructure-sensitive modeling of high cycle fatigue*. Int. J. Fatigue, 2010. **32**(3): p. 512-525. https://doi.org/10.1016/j.ijfatigue.2009.03.021.

[49] Shankar, V., et al., *Low cycle fatigue behavior and microstructural evolution of modified 9Cr–1Mo ferritic steel*. Mater. Sci. Eng. A, 2006. **437**(2): p. 413-422. https://doi.org/10.1016/j.msea.2006.07.146.

[50] Deb, K. and H. Jain, *An Evolutionary Many-Objective Optimization Algorithm Using Reference-Point-Based Nondominated Sorting Approach, Part I: Solving Problems With Box Constraints*. IEEE Transactions on Evolutionary Computation, 2014. **18**(4): p. 577-601. 10.1109/TEVC.2013.2281535.

[51] Jain, H. and K. Deb, *An Evolutionary Many-Objective Optimization Algorithm Using Reference-Point Based Nondominated Sorting Approach, Part II: Handling Constraints and Extending to an Adaptive Approach*. IEEE Transactions on Evolutionary Computation, 2014. **18**(4): p. 602-622. 10.1109/TEVC.2013.2281534.

[52] Sofinowski, K.A., et al., *Layer-wise engineering of grain orientation (LEGO) in laser powder bed fusion of stainless steel 316L.* Addit. Manuf., 2021. **38**: p. 101809. https://doi.org/10.1016/j.addma.2020.101809.

[53] Sofinowski, K., M. Wittwer, and M. Seita, *Encoding data into metal alloys using laser powder bed fusion.* Addit. Manuf., 2022. **52**: p. 102683. https://doi.org/10.1016/j.addma.2022.102683.

[54] Lu, H., et al., *Tailoring microstructure of additively manufactured Ti6Al4V titanium alloy using hybrid additive manufacturing technology.* Addit. Manuf., 2023: p. 103416.

[55] Quey, R., P. Dawson, and F. Barbe, *Large-scale 3D random polycrystals for the finite element method: Generation, meshing and remeshing.* Comput. Method App. M., 2011. **200**(17-20): p. 1729-1745.

[56] Quey, R. and L. Renversade, *Optimal polyhedral description of 3D polycrystals: Method and application to statistical and synchrotron X-ray diffraction data.* Comput. Method App. M., 2018. **330**: p. 308-333.

[57] Lütjering, G. and J.C. Williams, *Titanium matrix composites*. 2007: Springer.

[58] Hielscher, R., H. Schaeben, and H. Siemes, *Orientation Distribution Within a Single Hematite Crystal.* Math. Geosci., 2010. **42**(4): p. 359-375. https://doi.org/10.1007/s11004-010-9271-z.

[59] Niessen, F., et al., *Parent grain reconstruction from partially or fully transformed microstructures in MTEX.* J. Appl. Crystallogr., 2022. **55**(1): p. 180-194.

[60] Dawson, P.R. and D.E. Boyce, *FEpX--Finite element polycrystals: Theory, finite element formulation, numerical implementation and illustrative examples*, in *arXiv preprint arXiv:1504.03296*. 2015.

[61] Zhang, Y., et al., *Microstructure-Based Multiscale Modeling of Deformation in MarBN Steel under Uniaxial Tension: Experiments and Finite Element Simulations.* Materials, 2023. **16**(14): p. 5194. https://doi.org/10.3390/ma16145194.

[62] Li, R., et al., *A CPFEM based theoretical analysis of strains resolved by the microstructural feature tracking method.* IOP Conf. Ser.: Mater. Sci. Eng., 2022. **1249**(1): p. 012056. 10.1088/1757-899X/1249/1/012056.

[63] Somlo, K., et al., *Anisotropic tensile behaviour of additively manufactured Ti-6Al-4V simulated with crystal plasticity.* Mechanics of Materials, 2021. **162**: p. 104034. https://doi.org/10.1016/j.mechmat.2021.104034.

[64] Azhari, F., et al., *A Comparison of Statistically Equivalent and Realistic Microstructural Representative Volume Elements for Crystal Plasticity Models.* Integr. Mater. Manuf. I., 2022. **11**(2): p. 214-229. 10.1007/s40192-022-00257-4.

[65] Abadi, M., et al. *Tensorflow: a system for large-scale machine learning*. in *12th USENIX Symposium on Operating Systems Design and Implementation*. 2016. Savannah, GA, USA.

[66] LeCun, Y., et al., *Gradient-based learning applied to document recognition.* Proc. IEEE, 1998. **86**(11): p. 2278-2324.

[67] Goodfellow, I., et al., *Generative adversarial networks.* Commun. ACM, 2020. **63**(11): p. 139-144.

[68] Mirza, M. and S. Osindero, *Conditional generative adversarial nets.* arXiv preprint arXiv:1411.1784, 2014.

[69] Isola, P., et al. *Image-to-image translation with conditional adversarial networks*. in *Proceedings of the IEEE conference on computer vision and pattern recognition*. 2017.

[70] Yang, Z. and M.J. Buehler, *Fill in the Blank: Transferrable Deep Learning Approaches to Recover Missing Physical Field Information.* Advanced Materials, 2023. **35**(23): p. 2301449. https://doi.org/10.1002/adma.202301449.

[71] Ronneberger, O., P. Fischer, and T. Brox. *U-Net: Convolutional Networks for Biomedical Image Segmentation*. in *Medical Image Computing and Computer-Assisted Intervention – MICCAI 2015*. 2015. Cham: Springer International Publishing. https://doi.org/10.1007/978-3-319-24574-4.

[72] Springenberg, J.T., et al., *Striving for simplicity: The all convolutional net.* arXiv preprint arXiv:1412.6806, 2014.

[73] Radford, A., L. Metz, and S. Chintala, *Unsupervised representation learning with deep convolutional generative adversarial networks.* arXiv preprint arXiv:1511.06434, 2015.

[74] Gad, A.F., *Pygad: An intuitive genetic algorithm python library*, in *arXiv preprint arXiv:2106.06158*. 2021.

[75] Blank, J. and K. Deb, *Pymoo: Multi-Objective Optimization in Python.* IEEE Access, 2020. **8**: p. 89497-89509. 10.1109/ACCESS.2020.2990567.

[76] Fonseca, C.M., L. Paquete, and M. Lopez-Ibanez. *An Improved Dimension-Sweep Algorithm for the Hypervolume Indicator*. in *2006 IEEE International Conference on Evolutionary Computation*. 2006. https://doi.org/10.1109/CEC.2006.1688440.

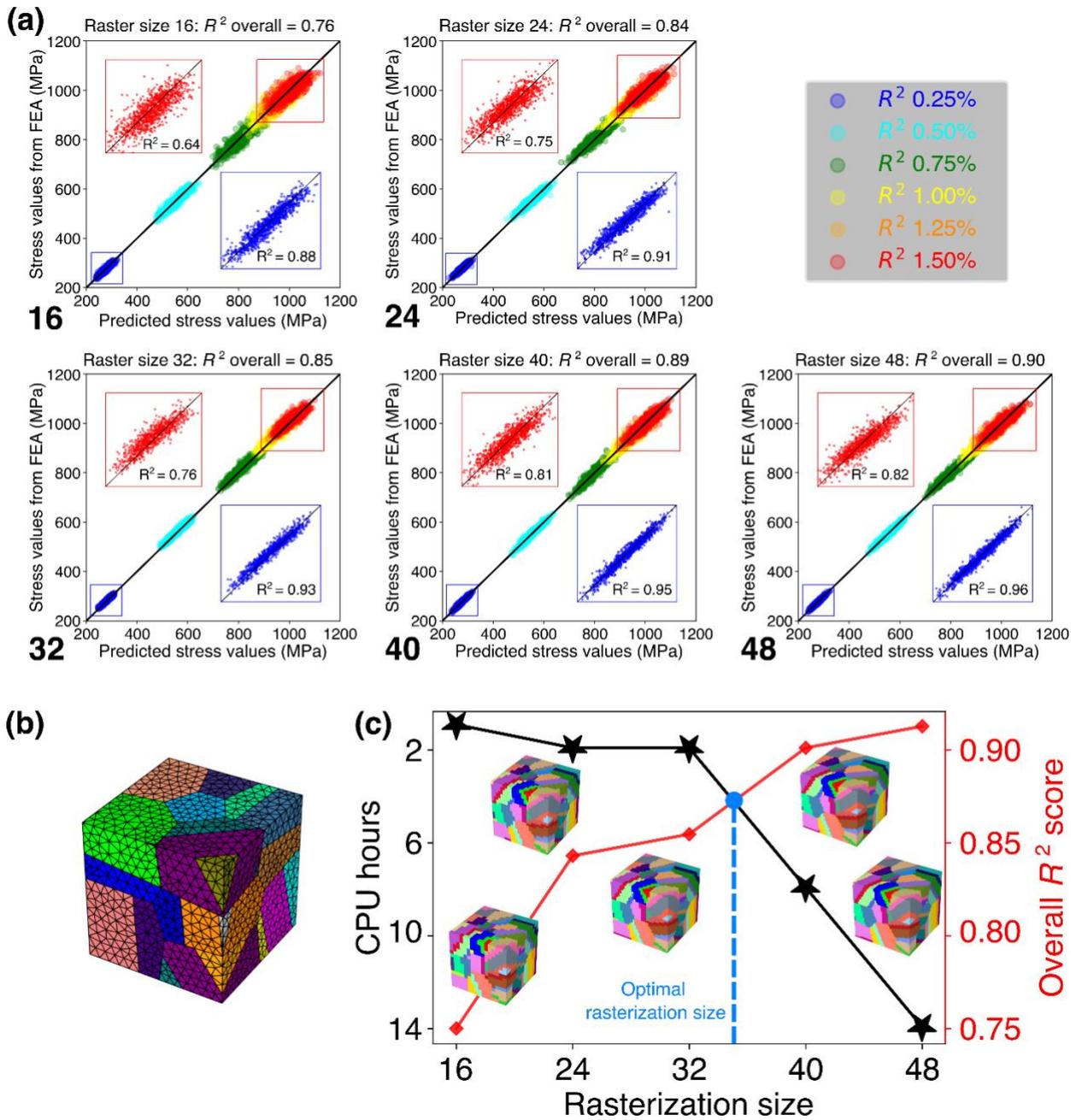

**Figure S1. Parametric study on rasterization levels and 3D-CNN model performances.** (**a**) Parity plots showing model performance with rasterization levels at 16, 24, 32, 40, and 48. The insets are zoomed-in views at 0.25% (blue) and 1.50% (red) strains, respectively. (**b**) A sample microstructure meshed with tetrahedral meshes used for FEA simulation. (**c**) Traded-off plot of CPU hours and overall R² score with changing rasterization size.

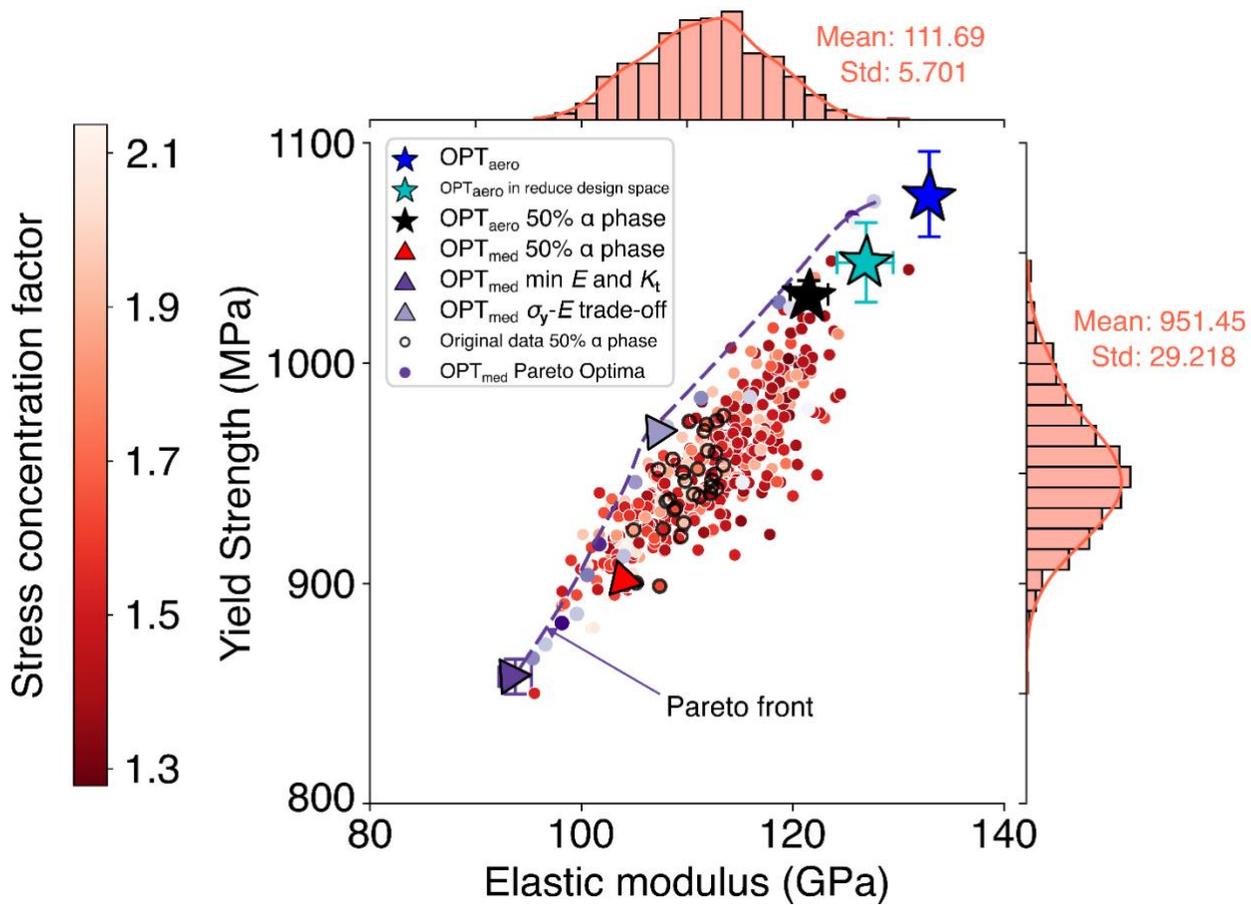

**Figure S2. Optimization results and original data distributions in 2D.** The red spectrum scattered points are datapoints from the original dataset, coloured by their corresponding stress concentration factor ($K_t$) values. The purple spectrum scattered points are the nondominated solutions identified in OPT$_{med1}$. Data points with 50% α phase from the original dataset are circled in black. Mechanical properties of the microstructures identified by optimizations are plotted with their means and standard deviations, except for the OPT$_{med}$ $\sigma_y$-$E$ trade-off case, for which the result from OPT$_{med1}$ is shown. Distribution histograms of the elastic modulus ($E$) and yield strengths ($\sigma_y$) are visualized on the top and right sides of the plot, respectively.

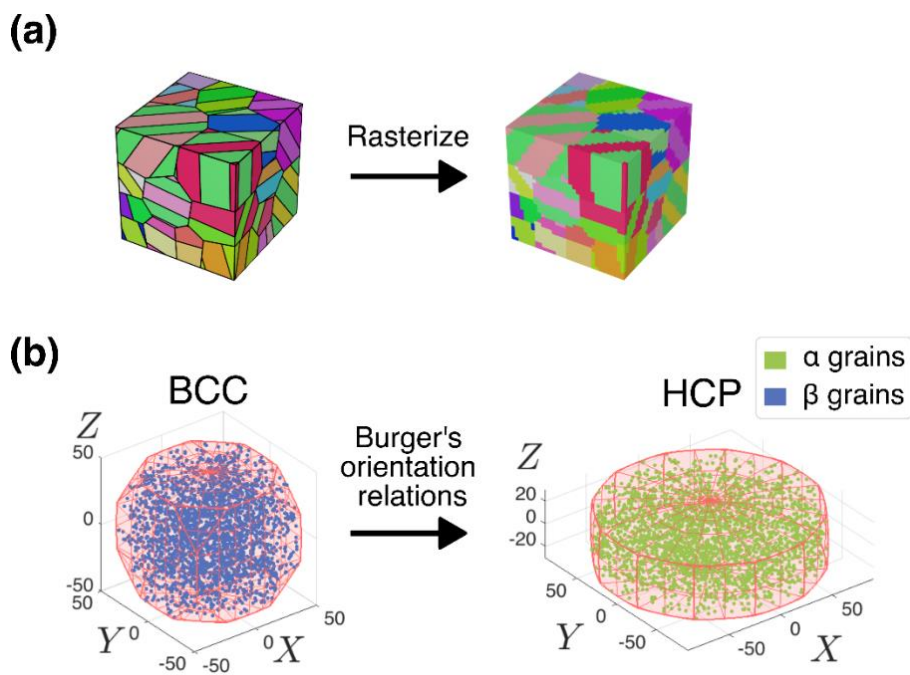

**Figure S3. Microstructure preparation.** (**a**) Rasterizing an artificially generated microstructure into a 32 x 32 x 32 3D image to use as deep learning model input. (**b**) Orientation distributions of the α and β grains in the dataset, plotted in their corresponding fundamental regions, respectively. The α grain orientations are calculated from the β grain orientations by enforcing the Burger's orientation relations.

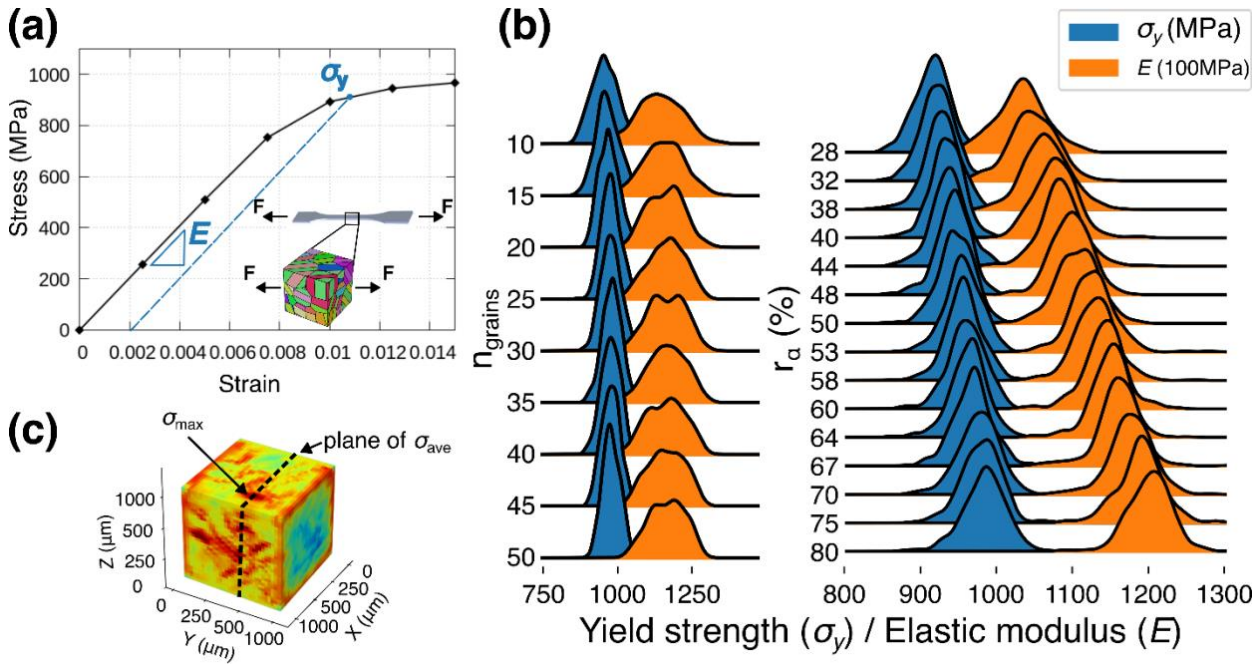

**Figure S4. Mechanical responses and the distribution of $\sigma_y$ and $E$.** (**a**) A sample stress-strain curve for a 3D microstructure under uniaxial tension with 0.2% offset yield strength ($\sigma_y$) and elastic modulus ($E$) marked. The inset displays a tensile test coupon under uniaxial loading and how a 3D microstructure represents its cross section in this work. (**b**) Distributions of $\sigma_y$ and $E$ for nearly 6,000 data in the dataset. The left shows the distributions' change with number of prior β grains ($n_{grains}$), while the right shows their change with α phase volume fraction ($r_a$). (**c**) Sample von Mises stress field with $\sigma_{max}$ and $\sigma_{ave}$ plane marked.

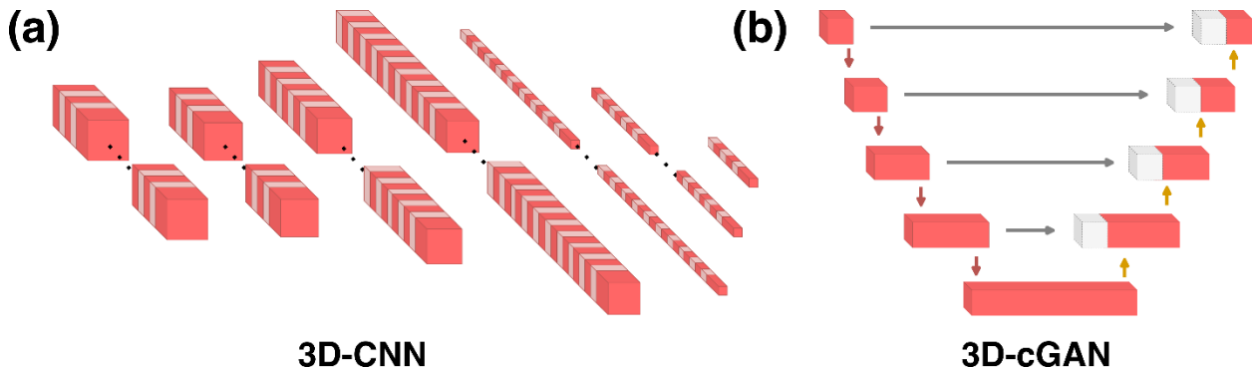

**(a)** **3D-CNN**

**(b)** **3D-cGAN**

**Figure S5. Schematic illustration of proposed 3D deep learning model architectures.** (**a**) 3D convolutional neural network (3D-CNN) model. (**b**) 3D conditional generative adversarial network (3D-cGAN) model.

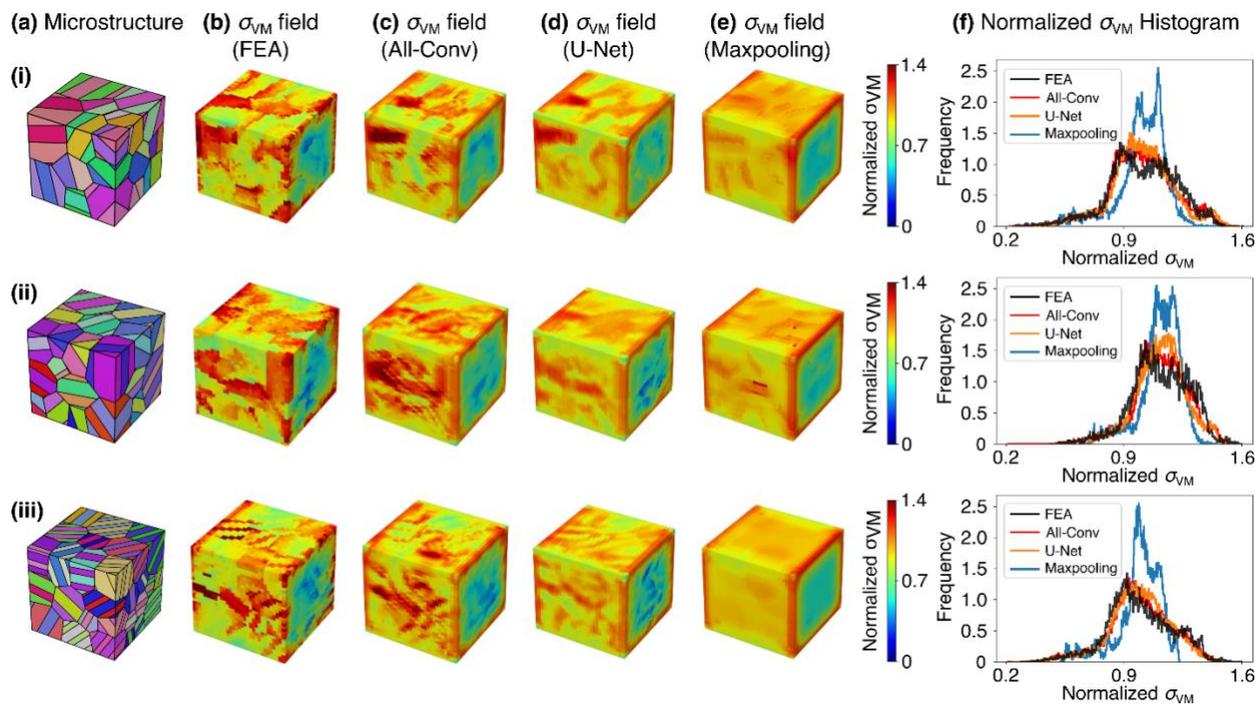

**Figure S6. Generator performance comparison on three randomly selected microstructures.** (**a**) Selected microstructures. (**b**) – (**e**) von Mises stress distribution fields by FEA calculation, all-convolutional generator prediction, original U-Net generator prediction, all-max-pooling generator prediction, respectively. (**f**) von Mises stress histograms of stress distributions from FEA calculation and various generators prediction.

**Table S1. Mean absolute errors and mean absolute percentage errors of the 3D-CNN model predictions for each by stress point and for the elastic and plastic regimes.**

| Strains (%) | 0.25 | 0.50 | 0.75 | 1.00 | 1.25 | 1.50 |
|---|---|---|---|---|---|---|
| Mean (MPa) | 280.13 | 558.66 | 815.39 | 944.55 | 989.20 | 1009.35 |
| Mean absolute error (MAE, MPa) | 3.011 | 6.073 | 8.914 | 10.933 | 11.698 | 11.880 |
| Mean absolute percentage error (MAPE, %) | 1.075 | 1.087 | 1.093 | 1.157 | 1.183 | 1.177 |
| MAPE by regime (%) | Elastic regime | | | Plastic regime | | |
| | 1.085 | | | 1.172 | | |

**Table S2. Detailed results of various optimization tasks in inverse exploration.**

| | $\sigma_y$ (MPa) | $E$ (GPa) | $K_t$ | $\alpha$ phase fraction |
|---|---|---|---|---|
| **OPT$_{aero1}$** | 1054.8 | 132.2 | 1.24 | 0.924 |
| **OPT$_{aero2}$** | 1084.4 | 132.8 | 1.30 | 0.864 |
| **OPT$_{aero3}$** | 1090.5 | 133.7 | 1.34 | 0.828 |
| **Mean** | 1076.56 | 132.89 | 1.29 | 0.872 |
| **Std** | 22.418 | 4.148 | 0.004 | 0.077 |
| **OPT$_{aero1}$ without $K_t$** | 1098.5 | 138.2 | - | 0.930 |
| **OPT$_{aero2}$ without $K_t$** | 1105.1 | 135.1 | - | 0.923 |
| **OPT$_{aero3}$ without $K_t$** | 1116.6 | 139.1 | - | 0.913 |
| **Mean** | 1106.71 | 137.45 | - | 0.922 |
| **Std** | 9.178 | 2.134 | - | 0.009 |
| **OPT$_{aero1}$, 50% $\alpha$ phase** | 1024.9 | 119.7 | 1.41 | 0.500 |
| **OPT$_{aero2}$, 50% $\alpha$ phase** | 1030.5 | 121.1 | 1.39 | 0.500 |
| **OPT$_{aero3}$, 50% $\alpha$ phase** | 1001.1 | 116.4 | 1.33 | 0.500 |
| **Mean** | 1018.83 | 119.09 | 1.38 | 0.500 |
| **Std** | 15.569 | 2.425 | 0.036 | - |
| **OPT$_{med1}$ min $E$ and $K_t$** | 854.7 | 92.7 | 1.55 | 0.321 |
| **OPT$_{med2}$ min $E$ and $K_t$** | 850.3 | 92.6 | 1.54 | 0.292 |
| **OPT$_{med3}$ min $E$ and $K_t$** | 866.0 | 95.4 | 1.48 | 0.319 |
| **Mean** | 857.00 | 93.57 | 1.52 | 0.311 |
| **Std** | 8.099 | 1.589 | 0.038 | 0.016 |
| **OPT$_{med1}$ $\sigma_y$-$E$ trade-off** | 971.1 | 108.1 | 1.40 | 0.323 |
| **OPT$_{med2}$ $\sigma_y$-$E$ trade-off** | 960.0 | 108.8 | 1.55 | 0.287 |
| **OPT$_{med3}$ $\sigma_y$-$E$ trade-off** | 953.5 | 106.0 | 1.59 | 0.324 |
| **Mean** | 961.53 | 107.63 | 1.513 | 0.311 |
| **Std** | 7.267 | 1.190 | 0.082 | 0.017 |
| **OPT$_{med1}$, 50% $\alpha$ phase** | 901.8 | 104 | 1.35 | 0.500 |
| **OPT$_{med2}$, 50% $\alpha$ phase** | 904.2 | 103 | 1.37 | 0.500 |
| **OPT$_{med3}$, 50% $\alpha$ phase** | 897.7 | 105.4 | 1.29 | 0.500 |
| **Mean** | 901.2333 | 104.1333 | 1.336667 | 0.500 |
| **Std** | 3.286842 | 1.205543 | 0.041633 | - |

**Table S3. Detailed results of OPT$_{\text{areo}}$ in the inverse exploration in a reduced design space with α phase fraction of 28%-80%.**

|  | $\sigma_y$ (MPa) | $E$ (GPa) | $K_t$ | α phase fraction |
|---|---|---|---|---|
| **OPT$_{\text{aero1}}$** | 1066.4 | 129.6 | 1.29 | 0.779 |
| **OPT$_{\text{aero2}}$** | 1030.7 | 124.2 | 1.27 | 0.731 |
| **OPT$_{\text{aero3}}$** | 1039.7 | 127.1 | 1.30 | 0.773 |
| **Mean** | 1045.61 | 126.94 | 1.289 | 0.761 |
| **Std** | 18.558 | 2.723 | 0.018 | 0.026 |

**Table S4. Statistics of $\sigma_y$, $E$, and $K_t$ of the microstructures with 25 prior-β grains in the original dataset.**

| α phase ratio | 28%-80% | | | 50% only | | |
|---|---|---|---|---|---|---|
| **Properties** | $\sigma_y$ (MPa) | $E$ (GPa) | $K_t$ | $\sigma_y$ (MPa) | $E$ (GPa) | $K_t$ |
| **Count** | 473 | 473 | 473 | 30 | 30 | 30 |
| **Mean** | 951.45 | 111.69 | 1.61 | 944.63 | 110.24 | 1.65 |
| **Std** | 29.218 | 5.701 | 0.212 | 19.495 | 2.350 | 0.195 |
| **Min** | 850.1 | 95.6 | 1.28 | 898.79 | 104.95 | 1.34 |
| **25%** | 931.5 | 107.8 | 1.45 | 934.99 | 108.69 | 1.49 |
| **50%** | 949.4 | 111.9 | 1.52 | 945.22 | 110.42 | 1.61 |
| **75%** | 969.0 | 115.5 | 1.81 | 955.73 | 112.30 | 1.79 |
| **Max** | 1046.4 | 130.9 | 2.14 | 976.18 | 113.48 | 2.07 |
| **Microstructure with maximum $\sigma_y$ and $E$** | 1046.4 | 123.64 | 1.40 | 976.18 | 113.48 | 1.60 |
| **Microstructure with minimum $E$** | 850.1 | 95.6 | 1.52 | 924.25 | 104.95 | 1.86 |

**Table S5. Material elastic constants used in FEA simulation.**

| Phases | $C_{11}$ (GPa) | $C_{12}$ (GPa) | $C_{13}$ (GPa) | $C_{44}$ (Gpa) |
|--------|----------------|----------------|----------------|----------------|
| **a** | 169.66 | 88.66 | 61.66 | 42.50 |
| **b** | 133.10 | 95.10 | - | 42.70 |

**Table S6. Initial Slip strengths and plasticity parameters used for both α and β phases in FEA simulation. The parameters shaded in light orange are assumed parameters.**

| $g_{0,b}$ (MPa) | $g_{0,b}$ (MPa) | $g_{0,p}$ (MPa) | $g_{0,p}$ (MPa) | $h_0$ (MPa) | $g_{s0}$ (MPa) | m | m' | $\dot{\gamma}_0$ $(s^{-1})$ | $\dot{\gamma}_{s0}$ $(s^{-1})$ |
|---|---|---|---|---|---|---|---|---|---|
| **390** | 468 | 390 | 663 | 190 | 530 | 0.01 | 0.01 | 1.0 | $5\times10^{10}$ |

**Table S7. 3D-CNN model architecture.**

| Layer # | Layer type* | Output shape |
|---|---|---|
| 1 | Input layer | 32x32x32x4 |
| 2 | Conv3D+BatchNormalization | 30x30x30x32 |
| 3 | Conv3D+BatchNormalization | 28x28x28x32 |
| 4 | Conv3D+BatchNormalization | 26x26x26x64 |
| 5 | Conv3D+BatchNormalization+GlobalAveragePooling3D | 128 |
| 6 | Dense+Dropout | 64 |
| 7 | Output layer | 6 |

*Layer type use the abbreviate terminology in TensorFlow.

**Table S8. 3D-cGAN model architecture.**

| Model type | Layer # | Layer type* | Output shape | Pre-trained | # of trainable parameters |
|---|---|---|---|---|---|
| **Generator (All convolutional)** | 1 | Input layer | 32x32x32x4 | - | 859,265 with transfer learning / 1,167,361 without transfer learning |
| | 2 | Conv3D+BatchNormalization | 30x30x30x32 | Y | |
| | 3 | Conv3D+BatchNormalization | 28x28x28x32 | Y | |
| | 4 | Conv3D+BatchNormalization | 26x26x26x64 | Y | |
| | 5 | Conv3D+BatchNormalization | 24x24x24x128 | Y | |
| | 6 | Conv3DTanspose+ BatchNormalization | 26x26x26x128 | Y | |
| | 7 | Conv3DTanspose+ BatchNormalization | 28x28x28x64 | N | |
| | 8 | Conv3DTanspose+ BatchNormalization | 30x30x30x32 | N | |
| | 9 | Output layer | 32x32x32x1 | N | |
| **Discriminator** | 1 | Input layer | 32x32x32x4 + 32x32x32x1 | - | |
| | 2 | Concatenate layer | 32x32x32x5 | | |
| | 3 | Conv3D | 30x30x30x32 | - | |
| | 4 | Conv3D+BatchNormalization | 28x28x28x32 | - | |
| | 5 | Conv3D+BatchNormalization | 26x26x26x64 | | |
| | 6 | ZeroPadding3D | 28x28x28x64 | - | |
| | 7 | Conv3D+BatchNormalization | 26x26x26x128 | | |
| | 8 | ZeroPadding3D | 28x28x28x128 | | |
| | 9 | Conv3D | 26x26x26x1 | | |

*Layer type use the abbreviate terminology in TensorFlow.

**Table S9. Architectures of 3D-cGAN model comparable generators.**

| Model type | Layer # | Layer type* | Output shape | # of trainable parameters | Transfer learning |
|---|---|---|---|---|---|
| **Original U-Net** | 1 | Input layer | 32x32x32x4 | 950,401 | No |
| | 2 | Conv3D+Maxpooling | 30x30x30x32 | | |
| | 3 | Conv3D+Maxpooling | 28x28x28x32 | | |
| | 4 | Conv3D+ Maxpooling | 26x26x26x64 | | |
| | 5 | Conv3D+ Maxpooling | 24x24x24x128 | | |
| | 6 | Conv3DTanspose+ BatchNormalization | 26x26x26x128 | | |
| | 7 | Conv3DTanspose+ BatchNormalization | 28x28x28x64 | | |
| | 8 | Conv3DTanspose+ BatchNormalization | 30x30x30x32 | | |
| | 9 | Output layer | 32x32x32x1 | | |
| **All-max-pooling architecture** | 1 | Input layer | 32x32x32x4 | 302,125 | No |
| | 2 | Maxpooling | 30x30x30x32 | | |
| | 3 | Maxpooling | 28x28x28x32 | | |
| | 4 | Maxpooling | 26x26x26x64 | | |
| | 5 | Maxpooling | 24x24x24x128 | | |
| | 6 | Conv3DTanspose+ BatchNormalization | 26x26x26x128 | | |
| | 7 | Conv3DTanspose+ BatchNormalization | 28x28x28x64 | | |
| | 8 | Conv3DTanspose+ BatchNormalization | 30x30x30x32 | | |
| | 9 | Output layer | 32x32x32x1 | | |

*Layer type use the abbreviate terminology in TensorFlow.